\documentclass[useAMS]{mn2e}
\usepackage{graphicx}
\usepackage{subfig}
\usepackage{amssymb}
\usepackage{natbib}
\usepackage{amsmath}
\usepackage{multirow} 
\usepackage{booktabs,threeparttable}
\usepackage{color}
\usepackage[colorlinks,citecolor=black]{hyperref}
\usepackage{graphicx,url,times}
\usepackage{array}
\usepackage{gensymb}
\usepackage{hyperref}
\usepackage{placeins}
\usepackage{dcolumn}
\DeclareGraphicsExtensions{.ps,.pdf,.png}
\makeatletter
\def\fps@figure{htbp}
\makeatother

\definecolor{grey}{rgb}{0.4,0.6,0.6}
\newcolumntype{L}{D{.}{.}{2,1}}

\vspace{-1cm}

\newcolumntype{P}[1]{>{\raggedleft\arraybackslash}p{#1}}

\begin{document}
\title[Galaxy Cluster Mass Reconstruction]{Galaxy Cluster Mass Reconstruction Project: \\
II. Quantifying scatter and bias using contrasting mock catalogues}
\author[Old et al.]{L. Old$^{1}$\thanks{E-mail:
ppxlo@nottingham.ac.uk}, R. Wojtak$^{2,3,4}$, G. A. Mamon$^{5}$ R. A. Skibba$^{6}$, F. R. Pearce$^{1}$,
D. Croton$^{7}$, S. Bamford$^{1}$,  \newauthor P. Behroozi$^{8}$, R. de Carvalho$^{9}$, J. C. Mu\~{n}oz-Cuartas$^{10}$, D. Gifford$^{11}$,  M. E. Gray$^{1}$, \newauthor A. von der Linden$^{2,3,12}$, M.R. Merrifield$^{1}$, S. I. Muldrew$^{13}$, 
V. M\"uller$^{14}$, R. J. Pearson$^{15}$, \newauthor T. J. Ponman$^{15}$,   E. Rozo$^{4, 16}$, E. Rykoff$^{4}$, A. Saro$^{17}$,
T. Sepp$^{18}$, C. Sif\'on$^{19}$ and E. Tempel$^{18}$ \\
$^{1}$School of Physics and Astronomy, University of Nottingham, Nottingham, NG7 2RD, UK\\ 
$^{2}$Dark Cosmology Centre, Niels Bohr Institute, University of Copenhagen, Juliane Maries Vej 30, DK-2100 Copenhagen, Denmark\\
$^{3}$Kavli Institute
for Particle Astrophysics and Cosmology, Stanford University, 452
Lomita Mall, Stanford, CA 94305-4085, USA\\ 
$^{4}$SLAC National Accelerator Laboratory, Menlo Park, CA 94025, USA\\
$^{5}$Institut
d\'Astrophysique de Paris (UMR 7095 CNRS $\&$ UPMC), 98 bis Bd Arago, F-75014 Paris, France\\
$^{6}$Center
for Astrophysics and Space Sciences, Department of Physics, University
of California, 9500 Gilman Dr, San Diego, CA 92093, USA\\ 
$^{7}$Centre
for Astrophysics \& Supercomputing, Swinburne University of
Technology, PO Box 218, Hawthorn, VIC 3122, Australia\\ 
$^{8}$Space Telescope Science Institute, Baltimore, MD 21218 USA\\
$^{9}$Instituto Nacional de Pesquisas Espaciais, MCT, S.J. Campos, Brazil\\ 
$^{10}$Group for Computational Physics and Astrophysics, Instituto de Fisica, Universidad de Antioquia, Medellin, Colombia\\
$^{11}$Department of Astronomy, University of Michigan, 500 Church
St. Ann Arbor, MI, USA\\
$^{12}$Department of Physics, Stanford University, 382 Via Pueblo Mall, Stanford, CA 94305-4060, USA\\
$^{13}$Department of Physics and Astronomy, University of Leicester, University Road, Leicester, LE1 7RH, UK\\ 
$^{14}$Leibniz-Institut f\"ur Astrophysik Potsdam, An der Sternwarte 16, D-14482 Potsdam, Germany\\ 
$^{15}$School
of Physics and Astronomy, University of Birmingham, Birmingham, B15
2TT, UK\\ 
$^{16}$Department of Physics, University of Arizona, Tucson, AZ, 85721\\
$^{17}$Department of Physics,
Ludwig-Maximilians-Universit\"{a}t, Scheinerstr. 1, D-81679
M\"{u}nchen, Germany\\ 
$^{18}$Tartu Observatory, Observatooriumi 1, 61602 T\~oravere,
Estonia\\ 
$^{19}$Leiden Observatory, Leiden University,
P.O. Box 9513, NL-2300 RA Leiden, The Netherland\\
\date{Accepted ??. Received ??; in original form ??}}
\pagerange{\pageref{firstpage}--\pageref{lastpage}} \pubyear{2014}
\maketitle

\label{firstpage}
\begin{abstract}
This article is the second in a series in which we perform an extensive comparison of various galaxy-based cluster mass estimation techniques that utilise the positions, velocities and colours of galaxies. Our aim is to quantify the scatter, systematic bias and completeness of cluster masses derived from a diverse set of 25 galaxy-based methods using two contrasting mock galaxy catalogues based on a sophisticated halo occupation model and a semi-analytic model.  
Analysing 968 clusters, we find a wide range in the RMS errors in $\log M_{\rm 200c}$ delivered by the different methods (0.18 to 1.08 dex,  i.e., a factor of $\sim$1.5 to 12), with abundance matching and richness methods providing the best results, irrespective of the input model assumptions. In addition, certain methods produce a significant number of catastrophic cases where the mass is under- or over-estimated by a factor greater than 10. Given the steeply falling high-mass end of the cluster mass function, we recommend that richness or abundance
matching-based methods are used in conjunction with these methods as a sanity check for studies selecting high mass clusters. We see a stronger correlation of the recovered to input number of galaxies for both catalogues in comparison with the group/cluster mass, however, this does not guarantee that the correct member galaxies are being selected. We do not observe significantly higher scatter for either mock galaxy catalogues. Our results have implications for cosmological analyses that utilise the masses, richnesses, or abundances of clusters, which have different uncertainties when different methods are used.
\end{abstract}


%----------------------------------------------------------------
\section{Introduction}
Statistical studies of the galaxy cluster population, in particular the cluster mass function, provide indispensable knowledge of cosmological model parameters (see \citealt{2011ARA&A..49..409A} for a review, \citealt{2012ApJ...745...16T}), large scale structure (e.g., \citealt{1988ARA&A..26..631B}; \citealt{2001AJ....122.2222E};
\citealt{2005MNRAS.357..608Y}; \citealt{2008ApJ...676..206P};
\citealt{2013MNRAS.430..134W}) and galaxy evolution (e.g., \citealt{2003MNRAS.346..601G};
\citealt{2005ApJ...623..721P}; \citealt{2008MNRAS.391..585M}). However, deducing accurate masses of these gravitationally bound structures remains a fundamental challenge for current and future cosmological studies.

A variety of techniques exist to detect galaxy clusters, and from this their masses can be estimated in a number of different ways.
However, cluster masses cannot be directly measured, but only indirectly inferred from observed properties that are correlated with mass. To maximise the constraining power of clusters for future cosmological surveys, it is essential to characterise the level of scatter and systematic bias associated with these mass proxies.  The \textit{Galaxy Cluster Mass Reconstruction Project} was created in order to ascertain how accurately we can measure cluster masses using techniques that rely upon the positions, velocities, colours and magnitudes of galaxies. Our goals are to quantify the systematic bias, intrinsic scatter and completeness that these methods produce and try to enhance their performance by deducing which type of method (or combination of methods) is best for any given observational set-up.

There are three general steps that galaxy-based techniques follow. The first is to locate the cluster overdensity and determine the cluster centre, the second is to choose which galaxies are members of the cluster and the final step is to use the properties of this membership to estimate a cluster mass. Popular cluster finding techniques include using Red Sequence filtering techniques (e.g., \citealt{2000AJ....120.2148G};
\citealt{2012MNRAS.420.1861M}; \citealt{2014ApJ...785..104R}) 
and brightest cluster galaxy (BCG) searches (e.g., \citealt{2005MNRAS.356.1293Y};
\citealt{2007ApJ...660..221K}). 
Friends-Of-Friends (FOF) group-finding algorithm based methods are also widely used (e.g., \citealt{2006ApJS..167....1B};
\citealt{2008AJ....135..809L}; \citealt{2014ApJ...788..109J};
 \citealt{Tempel+14}, see FOF optimisation study of \citealp{DuarteMamon14}), 
along with methods based upon Voronoi tessellation (e.g., \citealt{2002ApJ...580..122M};
\citealt{2004AJ....128.1017L}; \citealt{2009MNRAS.395.1845V};
\citealt{2011ApJ...727...45S}). Finally, the magnitudes and positions of galaxies are also used to search for over-densities via the matched filter algorithm (e.g.,
\citealt{1996AJ....111..615P}; \citealt{1999A&A...345..681O};
\citealt{1999ApJ...517...78K}; \citealt{2009ApJ...698.1221M}).

The second procedure of galaxy-based mass estimation is to deduce accurate galaxy membership. Initial membership can be chosen in a variety of ways. Some methods use the galaxies obtained during the first step of the cluster overdensity search via the FOF algorithm (e.g., \citealt{2005MNRAS.357..608Y}; \citealt{2007ApJ...671..153Y}; \citealt{2012MNRAS.423.1583M}; \citealt{Tempel+14}; Pearson et al. in preparation). Other commonly used methods are to select galaxies within a specified region of the colour--magnitude space (e.g., \citealt{2013ApJ...772...47S}) or in projected phase space (e.g., \citealt{2007MNRAS.379..867V}; \citealt{2009MNRAS.399..812W}; \citealt{2013MNRAS.429.3079M}; \citealt{2013ApJ...768L..32G}; \citealt{2013ApJ...772...25S}; Pearson et al. in preparation). Though these techniques generate an impression of which galaxies are associated with a cluster, deducing which galaxies are true members of the cluster is often problematic due to interloping galaxies. These interlopers are close to but not gravitationally bound to the cluster and their inclusion can lead to strong bias in velocity dispersion based mass estimates (e.g.,
\citealt{1983MNRAS.204...33L}; \citealt{1997NewA....2..119B};
\citealt{1997ApJ...485...39C}; \citealt{2006A&A...456...23B};
\citealt{2007A&A...466..437W}; \citealt{2010A&A...520A..30M}). To avoid the
inclusion of these interloper galaxies, often methods use a variety of
techniques such as iterative clipping (\citealt{1977ApJ...214..347Y}) or the
Gapper technique (\citealt{1990AJ....100...32B};
\citealt{1993ApJ...404...38G}) to reach convergence on cluster
properties. Alternatively, this interloper contamination can be modelled when
performing density fitting (e.g., \citealt{2007A&A...466..437W}).

The final and often deemed most important step of galaxy-based techniques is to use properties of the refined membership to estimate the cluster mass. One of more traditional methods is to apply the virial theorem to the projected phase space distribution of member galaxies
(e.g., \citealt{1937ApJ....86..217Z}; \citealt{1977ApJ...214..347Y};
\citealt{Evrard:2008vo}), maintaining the assumption that the cluster is in
virial equilibrium (and sometimes including the surface
term, see \citealt{1986AJ.....92.1248T}). Perhaps the simplest of approaches to measure the mass is to use richness: the number of galaxies associated with the cluster above a certain magnitude limit (e.g., \citealt{2003ApJ...585..215Y}). The distribution of galaxies in projected phase space is also used to estimate cluster mass, assuming that the cluster follows a Navarro, Frenk and White (NFW)
density profile (\citealt{1996ApJ...462..563N};
\citealt{1997ApJ...490..493N}). Finally, in the caustic technique, the escape velocity profile is identified in
projected phase space through an abrupt decrease in phase space density at
higher velocities, delivering a cluster mass (e.g., \citealt{1997ApJ...481..633D}; \citealt{1999MNRAS.309..610D};
\citealt{2013ApJ...768L..32G}).

In our first study (\citealt{2014MNRAS.441.1513O}, hereafter Paper~I), we set out to determine the simplest-case baseline by using a clean well defined data set based on a Halo Occupation Distribution (HOD), hereafter referred to as `HOD1'. This simple model delivers spherically symmetric clusters, idealised substructure, a strong richness correlation and isotropic, isothermal Maxwellian velocities. For this straight-forward test, we found that, above $\rm 10^{14} M_{\odot}$, recovered cluster masses are correlated with the true underlying cluster mass with scatter of typically a factor of two. However, below $\rm 10^{14} M_{\odot}$, the scatter rises and rapidly approaches an order of magnitude. We also found that richness-based approaches produced the lowest scatter, though it is not clear if this is due to the simplicity of the HOD1 model used.

Paper~I raised important questions: would a more complex and realistic input galaxy catalogue change the performance of the different classes of methods in extracting accurate cluster masses? Is the success of the richness-based methods caused by the simplicity of the HOD model used to generate the input galaxy catalogue?  To address these questions, we test the performance of 25 different galaxy-based methods by using two mock galaxy catalogues that are produced using more sophisticated, observationally realistic and, most importantly, \textit{contrasting} models. Using two distinct mock catalogues for this test not only allows us to evaluate how, or if, our results vary as a result of the model we use, but also allows us to explore how different prescriptions of populating galaxies impacts the efficacy of mock galaxy catalogues.  The ultimate goal of this project is not only to rank cluster mass methods but to gain insight into how we can improve both the cluster mass measurement techniques and generate more realistic mock galaxy catalogues.

The article is organised as follows. 
We describe the mock galaxy catalogue in Section~2, and the
mass reconstruction methods applied to this catalogue are briefly described in Section~3. In Section~4, we provide details of our analysis and present our results on cluster mass and membership comparisons in Section~5. We end with a discussion of our results and conclusions in Section 6. Throughout the article we adopt a Lambda cold dark matter ($\Lambda$CDM) cosmology with $\Omega_{\rm 0}=0.27$, $\Omega_{\rm \Lambda}=0.73$, $\sigma_{\rm
 8}=0.82$ and a Hubble constant of $H_{\rm 0} =
100\,\rm{km\,s^{-1}}\,\rm{Mpc^{-1}}$ where $h=0.7$, although none of the conclusions depend strongly on these parameters.
%----------------------------------------------------------------
\section{Data}
\label{sec:Sim Data}
This article is the second in a series in which we perform an extensive comparison of galaxy-based techniques using two different mock galaxy catalogues. The two catalogues are produced by populating the underlying dark matter simulation with two sophisticated models that, importantly, are fundamentally different in nature. Both the more sophisticated HOD, referred to as `HOD2', and Semi-Analytic model, referred to as `SAM2', are described below along with a description of how the light cone was constructed. To deliver a sample containing both high mass clusters and lower mass groups, 1000 groups/clusters are selected from the two diverse mock galaxy catalogues by taking the 800 most massive and then the next 200 richest clusters. Any duplicate clusters present due to the way in which the light cones are constructed as well as clusters lying close to the edge of the cone are removed from the main analysis leaving, 968 groups/clusters.
%----------------------------------------------------------------
\subsection{Underlying dark matter simulation}
\label{sec:DM data}
We begin by using the Bolshoi dissipationless cosmological simulation which follows the evolution of $2048^{3}$ dark matter particles of mass $1.35 \times
10^{8}\,h^{-1} {\rm M_{\rm \odot}}$ from $z=80$ to $z=0$ within a box of side length $250\,h^{-1} {\rm Mpc}$ (\citealt{2011ApJ...740..102K}). The force resolution of the simulation is $1\,h^{-1}$ kpc and the halo catalogues are complete for haloes with circular velocity $V_{\rm circ}>50\,\rm{km\,s^{-1}}$ (corresponding to $M_{\rm 360\rho} \approx 1.5 \times
10^{10}\,h^{-1} {\rm M_{\rm \odot}}$). The simulation
adopts a flat $\Lambda$CDM cosmology with the following parameters:
$\Omega_{\rm 0}=0.27$, $\Omega_{\rm \Lambda}=0.73$, $\sigma_{\rm
 8}=0.82$, $n=0.95$ and $h=0.70$ and was run with the \textsc{ART} adaptive mesh refinement code. Dark matter haloes, substructure and tidal features are
identified using \textsc{ROCKSTAR}, a 6D FOF group-finder based on adaptive
hierarchical refinement (\citealt{2013ApJ...762..109B}). This halo finder has
been shown to recover halo properties with high accuracy and produces
consistent results with other halo finders
(\citealt{2011MNRAS.415.2293K}). The haloes and subhaloes found using
\textsc{ROCKSTAR} are then joined into hierarchical merging trees that
describe in detail how structures grow as the universe evolves.
%----------------------------------------------------------------
\subsection{Light cone construction}
The light cones used in this work were produced using the Theoretical
Astrophysical Observatory
(TAO\footnote{\hyperref[https://tao.asvo.org.au/tao/]{https://tao.asvo.org.au/tao/}},
\citealt{2014arXiv1403.5270B}), an online eResearch tool that provides access
to semi-analytic galaxy formation models and N-body simulations, including
tools which modify them to produce more realistic mock catalogues. 
%For our work 
Here, we use the light cone generation tool that remaps the original
spatial and temporal positions of each galaxy in the box onto an observer
cone specified by the user, which in our case subtends 60$^{\circ}$ by 60$^{\circ}$
on the sky, covering a redshift range of $ 0 < z < 0.15$. Note that
this cone is not flux-limited. However, as in Paper~I, we specify a minimum
$r$-band luminosity for the galaxies of $M_{r} = -19 + 5\log h$ for both the
HOD2 and SAM2 catalogues.
%----------------------------------------------------------------
\subsection{Halo Occupation Distribution model}
For the HOD2 model, a galaxy group/cluster catalogue was constructed with the
halo catalogue using an updated version of the model described in
\citet{2006MNRAS.369...68S} and \citet{2009MNRAS.392.1080S}.  We refer the
reader to these articles and Paper~I
for details.  Briefly, haloes are populated with galaxies whose luminosities
and colours are modelled such that they approximately reproduce the
luminosity function, colour-magnitude distribution, and luminosity- and
colour-dependant redshift- and real-space clustering in the Sloan Digital Sky
Survey (SDSS,\citealt{2000AJ....120.1579Y}).  An important assumption in this
HOD2 model is that all galaxy properties --their abundances, spatial
distributions, velocities, luminosities, and colours--are determined by parent halo mass alone, using the mass ($M_{\rm 200c}$) given by the ROCKSTAR algorithm.  
%While this model and its variants have considerable successes, if ``assembly
%bias" (the dependence of the spatial distribution of dark
%matter haloes upon properties other than mass) is a significant effect, then this assumption is an oversimplification
%(see \citealt{2014MNRAS.443.3044Z}; \citealt{2014arXiv1404.6524H}). For
%example, in their dynamical mass analysis of SDSS galaxies traced by their
%satellites, \cite{2013MNRAS.428.2407W} find that the dark matter halo
%concentrations of red galaxies are significantly greater than those of blue
%galaxies of the same halo mass.

The relevant model updates include the following.  First, the
\citet{2009MNRAS.392.1080S} model is extended by allowing for a dependence of the colour
distribution on halo mass at fixed luminosity (\citealt{2011MNRAS.410..210M};
\citealt{2013MNRAS.435.1313H}; \citealt{2013ApJ...767...92R}), and we include
colour gradients within haloes (\citealt{2009ApJ...699.1333H};
\citealt{2008MNRAS.387...79V}), which results in red galaxies having higher
number density concentrations than blue galaxies in haloes of a given
mass (as measured by \citealt{CollisterLahav05}). 
We include stellar masses based on the \citet{2009MNRAS.400.1181Z}
calibration, and the resulting distributions are approximately consistent
with the \citet{2013ApJ...767...50M} stellar mass function. Secondly, we update the concentration-mass relation and scatter by adopting those of \citet{2013MNRAS.428.2407W} and account for the fact that galaxies and subhaloes are less concentrated than dark matter (e.g., \citealt{2005ApJ...633..122H}; \citealt{2005MNRAS.357..608Y}, \citealt{2013MNRAS.428.2407W}) by adopting concentration index $c_\mathrm{gal}=c_\mathrm{DM}/1.5$.
Thirdly, the updated model includes a
treatment of dynamically unrelaxed systems, including some non-central
brightest halo galaxies, central galaxy velocity bias, and massive
substructures, all of which depend on host halo mass (see
\citealt{2011MNRAS.410..417S}; \citealt{2011MNRAS.416.2388S}).

With these changes, the motions of galaxies in the haloes are no longer
isothermal and isotropic, contrary to the HOD model used in Paper~I. For
haloes without these effects, the velocity dispersion profiles are isothermal
and isotropic as in Paper~I, with a velocity dispersion that depends on halo mass and
radius through the scaling $\sigma_\mathrm{\rm 200}^2=\tfrac{1}{2} G M_{\rm 200}/R_{\rm 200}$
(but see \citealt{2010A&A...520A..30M}; \citealt{2013MNRAS.430.2638M};
\citealt{2013MNRAS.434.2606O}).   
The updated model, including a more realistic velocity dispersion profile and
an anisotropy model, will be described in Skibba (in preparation). 
%----------------------------------------------------------------
\subsection{Semi-analytic model}
The Semi-Analytic Galaxy Evolution (SAGE) galaxy formation model used in this work (Croton et al. in preparation) is an updated version of that described in \cite{2006MNRAS.365...11C}. The merger trees described in Section~\ref{sec:DM data} form the backbone on which the model of galaxy formation is applied.
Inside each tree and at each redshift, virialised dark matter haloes are assumed to attract pristine gas from the surrounding environment, from which galaxies form and evolve.  The model tracks a wide range of galaxy formation physics, including reionisation of the inter-galactic medium at early times, the infall of this gas into haloes, radiative cooling of hot gas and the formation of cooling flows, star formation in the cold disk of galaxies and the resulting supernova feedback, black hole growth and active galactic nuclei (AGN) feedback through the `quasar' and `radio' epochs of AGN evolution, metal enrichment of the inter-galactic and intra-cluster medium from star formation, and galaxy morphology shaped through secular processes,  mergers and merger induced starbursts.

Each group identified by \textsc{ROCKSTAR} has one `central' galaxy whose central position and velocity is determined by averaging the positions and velocities of the subset of halo particles. Each group also has a number of `satellite' galaxies that maintain the positions and velocities of the subhaloes that merged with the parent halo. As a result, galaxies retain the `memory' of the dynamical history of the underlying DM simulation (this is not the case for the HOD2 model).

The SAGE model is primarily calibrated against $z=0$ observations, including the stellar mass function and SDSS-band luminosity functions, baryonic Tully-Fisher relation, black hole-bulge relation, and metallicity-stellar mass relation. At higher redshift the model provides a good match to the star formation rate density evolution and stellar mass function evolution. See \citet{2014ApJ...795..123L} and Croton et al. 2014 (in preparation) for more details and focused comparisons. 
%One would expect galaxy positions and velocities to be more realistic in the SAM2 catalogue, while galaxy colours may be more realistic in the HOD2 catalogue.
%----------------------------------------------------------------
\section{Mass Reconstruction Methods}
\label{sec:Mass Reconstruction Methods}
\begin{table*} 
 \caption{Summary of the cluster mass estimation
 methods. Listed is an acronym identifying the method, an
 indication of the main property used to undertake member galaxy selection
 and an indication of the method used to convert this membership
 list to a mass estimate. The type of observational data required as input for each method is listed in the fourth column. Note that acronyms denoted with an asterisk
 indicate that the method did not use our initial object
 target list but rather matched these locations at the end of their analysis. Please see Tables
 \ref{table:appendix_table_1} and \ref{table:appendix_table_2} in
 the appendix for more details on each method.} 
\begin{center} 
\tabcolsep 0.15cm
\begin{tabular}{l l l l l}
\hline
\multicolumn{1}{c}{Method}&Initial Galaxy Selection&Mass Estimation&Type of data required&Reference \\ \hline
PCN&Phase space&Richness&Spectroscopy&Pearson et al. (in prep.)\\
PFN*&FOF&Richness&Spectroscopy&Pearson et al. (in prep.)\\
NUM&Phase space&Richness&Spectroscopy&Mamon et al. (in prep.)\\
RM1&Red Sequence&Richness&Multi-band photometry, sample of central spectra&{\citet{2014ApJ...785..104R}}\\ 
RM2*&Red Sequence&Richness&Multi-band photometry, sample of central spectra&{\citet{2014ApJ...785..104R}}\\ 
ESC&Phase space& Phase space&Spectroscopy&{\citet{2013ApJ...768L..32G}}\\ 
MPO&Phase space& Phase space&Multi-band photometry, spectroscopy&{\citet{2013MNRAS.429.3079M}}\\ 
MP1& Phase space& Phase space&Spectroscopy&{\citet{2013MNRAS.429.3079M}}\\
RW& Phase space& Phase space&Spectroscopy&{\citet{2009MNRAS.399..812W}}\\
TAR*&FOF& Phase space&Spectroscopy&{\citet{Tempel+14}} \\
PCO& Phase space&Radius&Spectroscopy&Pearson et al. (in prep.)\\ 
PFO*&FOF& Radius& Spectroscopy&Pearson et al. (in prep.)\\
PCR& Phase space&Radius&Spectroscopy&Pearson et al. (in prep.)\\ 
PFR*&FOF&Radius&Spectroscopy&Pearson et al. (in prep.)\\
MVM*&FOF&Abundance matching&Spectroscopy&{\citet{2012MNRAS.423.1583M}}\\
AS1&Red Sequence&Velocity dispersion&Spectroscopy&{\citet{2013ApJ...772...47S}}\\
AS2&Red Sequence&Velocity dispersion&Spectroscopy&{\citet{2013ApJ...772...47S}}\\
AvL& Phase space&Velocity dispersion&Spectroscopy&{\citet{2007MNRAS.379..867V}}\\ 
CLE& Phase space&Velocity dispersion&Spectroscopy&{\citet{2013MNRAS.429.3079M}}\\ 
CLN& Phase space&Velocity dispersion&Spectroscopy&{\citet{2013MNRAS.429.3079M}}\\
SG1& Phase space&Velocity dispersion&Spectroscopy&{\citet{2013ApJ...772...25S}}\\
SG2& Phase space&Velocity dispersion&Spectroscopy&{\citet{2013ApJ...772...25S}}\\
SG3& Phase space&Velocity dispersion&Spectroscopy&{\citet{2009MNRAS.392..135L}}\\
PCS& Phase space&Velocity dispersion&Spectroscopy&Pearson et al. (in prep.)\\ 
PFS*&FOF&Velocity dispersion&Spectroscopy&Pearson et al. (in prep.)\\ 
\hline 
\end{tabular}
\end{center}
\label{table:basic_method_characteristics}
\end{table*} 
We present details of the additional cluster mass reconstruction methods tested in the second phase of the project and we highlight below any changes to the methods that participated in Phase I of the project. The type of data the methods require as input and a summary of the basic properties of all methods are listed in Table~\ref{table:basic_method_characteristics}. As for Phase I, we provide a more detailed overview of the methods in Tables~\ref{table:appendix_table_1} and \ref{table:appendix_table_2} in the appendix. Each method is identified by an acronym and the subsection titles for each method are given in the form (author name; initial galaxy selection technique, mass estimation property). The initial cluster membership is performed in three classes: projected phase space, FOF, or Red Sequence. The subsequent mass estimation is performed according to five classes of methods: richness, projected phase space, radii, velocity dispersions, or abundance matching. For detailed descriptions of methods that also participated in Phase I of the project, please refer to Paper~I.
\subsection{NUM (Mamon; Phase space, Richness)}
In Paper~I, NUM was based on the mass derived from
a robust linear fit to mock clusters analysed by the CLE mass estimation method, $\log \left (M_{\rm CLE}/{\rm M}_\odot\right) = a + b \log N_{1\,\rm Mpc,
  1333\,km\,s^{-1}}^{\rm CLE}$, which yielded $a = 12.02$ and $b=1.38$. 
Now NUM uses the robust bilinear fit to $\log \left (M_{\rm CLE}/{\rm
  M}_\odot\right) = a + b\,\log N_{1\,\rm Mpc, 1333\,km\,s^{-1}}^{\rm CLE} +
c \log (1+z)$. The metric radius is now 1 Mpc \emph{comoving}.
The constants are now $a=12.43, b=1.22, c=-4.25$ and
$a=12.21, b=1.24, c=-2.53$ for the HOD2 and SAM2 catalogues respectively.  
For a given richness $\log N_{1\,\rm Mpc,  1333\,km\,s^{-1}}^{\rm CLE}$, the
SAM2 masses are typically 0.19~dex lower than the HOD2 masses.

\subsection{RM1 (Rykoff \& Rozo; Red Sequence, Richness)}
The Red Sequence Matched-filter Probabilistic Percolation (redMaPPer)
algorithm \citep{2014ApJ...785..104R}, based on the optimised richness estimator $\lambda$ \citep{rykoffetal12}, is a photometric cluster finder that
identifies galaxy clusters as overdensities of Red Sequence galaxies. It has
excellent photo-$z$ performance and $\lambda$ has been found to be a low-scatter
mass proxy \citep{2014ApJ...783...80R,2014ApJ...785..104R}.  The algorithm is divided into two
stages: the first is a calibration stage where the Red Sequence model is
derived directly from the data, and the second is the cluster-finding stage.
Given a list of cluster/halo positions and estimated redshifts, it is also
possible to directly compute the richness and photo-$z$ given the Red Sequence
calibration.

As redMaPPer works entirely in observed magnitude space, all absolute
magnitudes from the input catalog are de-k-corrected to observed $g$ and $r$
magnitudes using {\tt sdss\_kcorrect} \citep{blanton07}.
Although redMaPPer can be run using multi-band data, the current data set
comprised only two bands.  We use a random sample of halo centres 
as the ``seed" spectroscopic galaxies used to calibrate the Red Sequence 
over the redshift range $0.05<z<0.15$.  Only the Phase 2 HOD2 galaxy sample
could be used with redMaPPer, as the colour model in the Phase 2 SAM2 does not
result in a prominent enough Red Sequence.

Two separate runs of redMaPPer are performed, with both using the same
calibration as described above.  The first run, denoted RM1, directly computed
the richness and photo-$z$ at the location of each halo (using the
true redshift as a starting point).  Mass estimates are made using
the abundance-matching estimate for $\lambda$-$M_{\rm 200c}$, calibrated by \citet{rykoffetal12},  which follows a power law (see Appendix B of
\citealt{rykoffetal12}). 
%----------------------------------------------------------------
\subsection{RM2 (Rykoff \& Rozo; Red Sequence, Richness)}
This method is similar to RM1, but is a full cluster finding
run using the algorithm of \citet{2014ApJ...785..104R}.  After detection, the clusters
are sorted by descending richness, and each cluster is matched to the nearest
halo within $3\sigma_{\rm z}$.  Mass estimates are performed as for RM1.  We note
that only for the RM2 run will there be any offsets between the redMaPPer
cluster centres and the halo centres.
%----------------------------------------------------------------
\subsection{SG3 (de Carvalho; Phase space, Velocity dispersion)}
SG3 is a method for the rejection of velocity interlopers to produce a final
list of cluster members, making no hypotheses about the dynamical status of
the cluster (e.g. \citealt{2007A&A...466..437W}). The algorithm is similar to
the one proposed by \citet{1996ApJ...473..670F} and used by SG1 and SG2. It
applies the Gapper technique in radial bins with sizes of $0.42 h^{-1}
\rm{Mpc}$ or larger, to guarantee at least 15 galaxies per bin. The procedure
is repeated until there are no more interlopers and the list of members is
used to estimate cluster properties. We perform virial analysis in an
analogous way to \citet{1998ApJ...505...74G}, \citet{2005A&A...433..431P},
\citet{2006A&A...456...23B} and \citet{2007A&A...461..397P}. First, we
compute the robust aperture velocity dispersion ($\sigma_{\rm ap}$) of the
cluster depending on the number of members available: gapper ($<$ 15) or
bi-weight ($\geq$15, \citealt{1990AJ....100...32B}). Then, $\sigma_{\rm ap}$
is corrected for redshift errors (\citealt{1980A&A....82..322D}) and an
estimate of 
% projected
virial radius is obtained following \citet{1998ApJ...505...74G}. These steps lead us to an initial virial mass estimate (equation 5 of \citealt{1998ApJ...505...74G}), which is then corrected for the surface pressure term (\citealt{1986AJ.....92.1248T}).

After applying such a correction, $R_{\rm 200c}$  is estimated considering the virial mass density. If $M_{V}$ is the virial mass in a volume of radius $R_{\rm A}$, then $R_{\rm 200c}$  = $R_{\rm A}\{ {\rho_{\rm V} /[200\rho_{\rm c}(z)]}\}^{1/2.4}$, where $\rho_{\rm V}$= 3$M_{\rm V}$/(4$\pi R^{3}_{\rm A}$) and $\rho_{\rm c}$(z) is the critical density at redshift z. Finally, assuming an NFW profile, we obtain $M_{\rm 200c}$ from the interpolation (most cases) or extrapolation of the virial mass $M_{\rm V}$ from $R_{\rm A}$ to $R_{\rm 200c}$ .  This procedure is analogous to what is done by  \citet{2006A&A...456...23B} and \citet{2007A&A...461..397P}.
%----------------------------------------------------------------
\subsection{MVM (M\"uller; FOF, Abundance Matching)} 
The member galaxy selection stage of MVM has been modified. In Phase I, an ellipsoidal boundary was used to define group/cluster membership. Now membership is determined by including all galaxies within a joint line-of-sight and plane-of-sky distance from the group centres until the background galaxy density is reached.
%----------------------------------------------------------------
\subsection{CLE \& CLN (Mamon; Phase space, Velocity dispersion)}
CLE and CLN are as in Paper~I, except that when the group is split by the gapper technique, the subsample containing the mean halo velocity is kept (instead of the largest one, as in Paper~I).
%\renewcommand{\tabcolsep}{0.3cm}
%\begin{table} 
% \caption{The type of observational data required as input for each method. Note that acronyms denoted with an asterisk
% indicate that the method did not use our initial object
% target list but rather matched these locations at the end of their analysis.} 
%\begin{center} 
%\tabcolsep 0.3cm
%\begin{tabular}{l l}
%\hline
%\multicolumn{1}{c}{Method}&Type of data required \\ \hline
%PCN&Spectroscopy\\
%PFN*&Spectroscopy\\
%NUM&Spectroscopy\\
%RM1& Multi-band photometry, sample of central spectra\\ 
%RM2*& Multi-band photometry, sample of central spectra\\ 
%ESC&Spectroscopy\\ 
%MPO&Multi-band photometry, spectroscopy\\ 
%MP1&Spectroscopy\\
%RW&Spectroscopy\\
%TAR*&Spectroscopy \\
%PCO&Spectroscopy\\ 
%PFO*&Spectroscopy\\
%PCR& Spectroscopy\\ 
%PFR*&Spectroscopy\\
%MVM*& Spectroscopy\\
%AS1&Spectroscopy\\
%AS2&Spectroscopy\\
%AvL& Spectroscopy\\ 
%CLE&Spectroscopy\\ 
%CLN&Spectroscopy\\
%SG1&Spectroscopy\\
%SG2&Spectroscopy\\
%SG3&Spectroscopy\\
%PCS&Spectroscopy\\ 
%PFS*&Spectroscopy\\
%\hline
%\end{tabular}
%\end{center}
%\label{table:data_methods_require}
%\end{table} 
%----------------------------------------------------------------
\section{Analysis}
\label{sec:analysis}
\begin{figure}
 \centering
 \includegraphics[trim = 0mm 0mm 0mm 0mm, clip, width=0.45\textwidth]{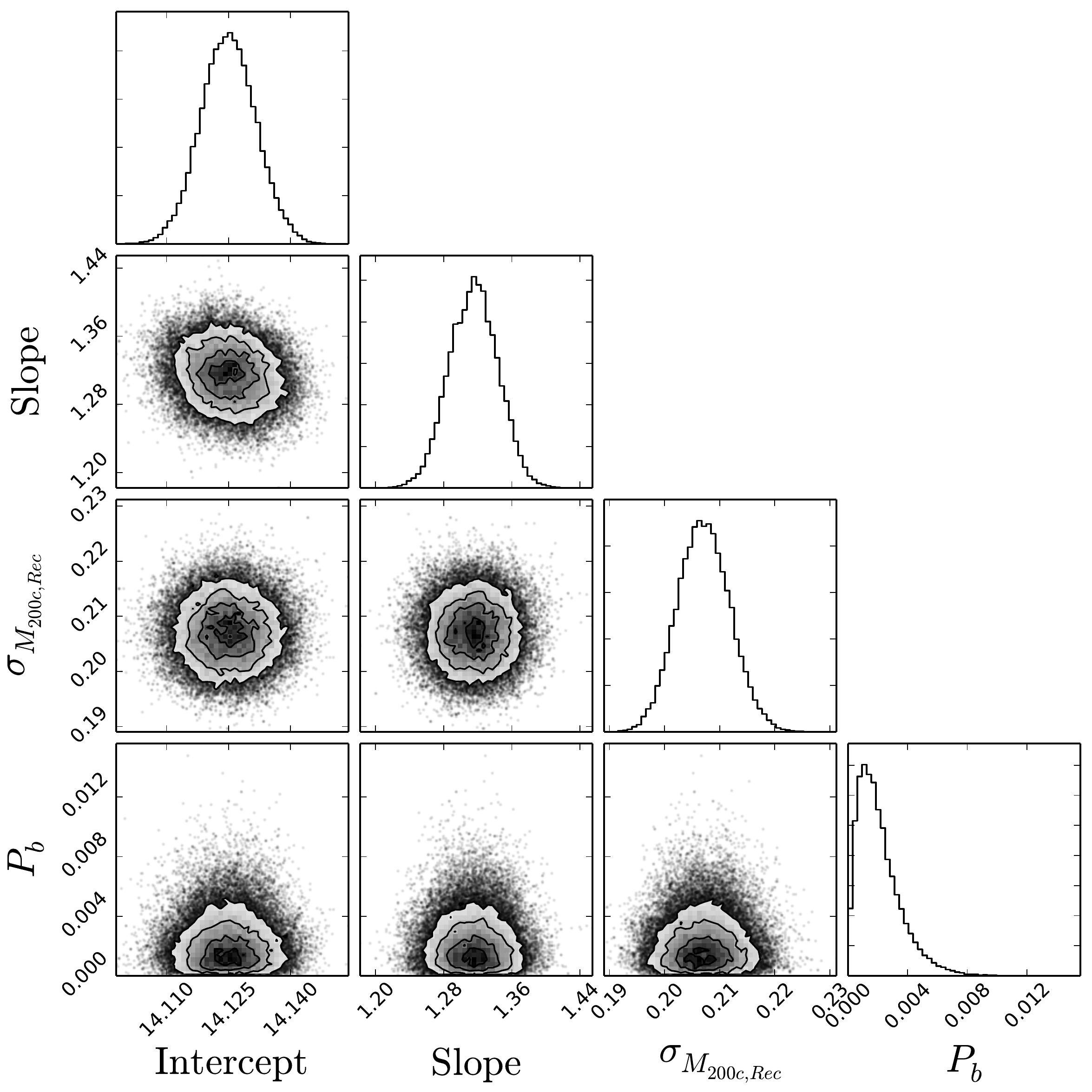}
 \caption{Example of the marginalised probability distributions produced by our MCMC
   analysis of the PCN method using the HOD2 catalogue for the parameters: the
   slope ($m$), the intercept ($c$), the scatter in the recovered mass
   ($\sigma_{M_{\rm Rec}}$) and the posterior fraction of data points belong
   to the `bad' outlier distribution ($P_{\rm b}$).
}
\label{fig:SAM_PCN_MCMC}
\end{figure}
We employ various statistics to examine the performance of the mass reconstruction methods including the Root Mean Square (RMS) difference between the recovered and input log mass, the scatter in the recovered mass, $\sigma_{M_{\rm Rec}}$, the scatter about the true mass $\sigma_{M_{\rm True}}$ and the bias. For the latter three statistics, we assume a model where there is a linear relationship between the recovered and true log mass and residual offsets in the recovered mass are drawn from a normal distribution. Instead of clipping outliers, we try the preferable approach of modelling the uncertainties in the data as justified in \citet{2010arXiv1008.4686H}. We take a Bayesian approach, computing a likelihood that is a sum of the probability of obtaining the data point assuming it is drawn from a `good' distribution and the probability of obtaining the data point assuming it is drawn from a `bad' outlier distribution. This ensures that the measured scatter is not affected by a very small number of extreme outliers. For example, in the case of a method that produces very low scatter in general but has, say, one or two extreme outliers, the measured scatter will not be falsely inflated. 
\begin{figure*}
 \centering
\includegraphics[trim = 45mm 52mm 55mm 57mm, clip, width=0.75\textwidth]{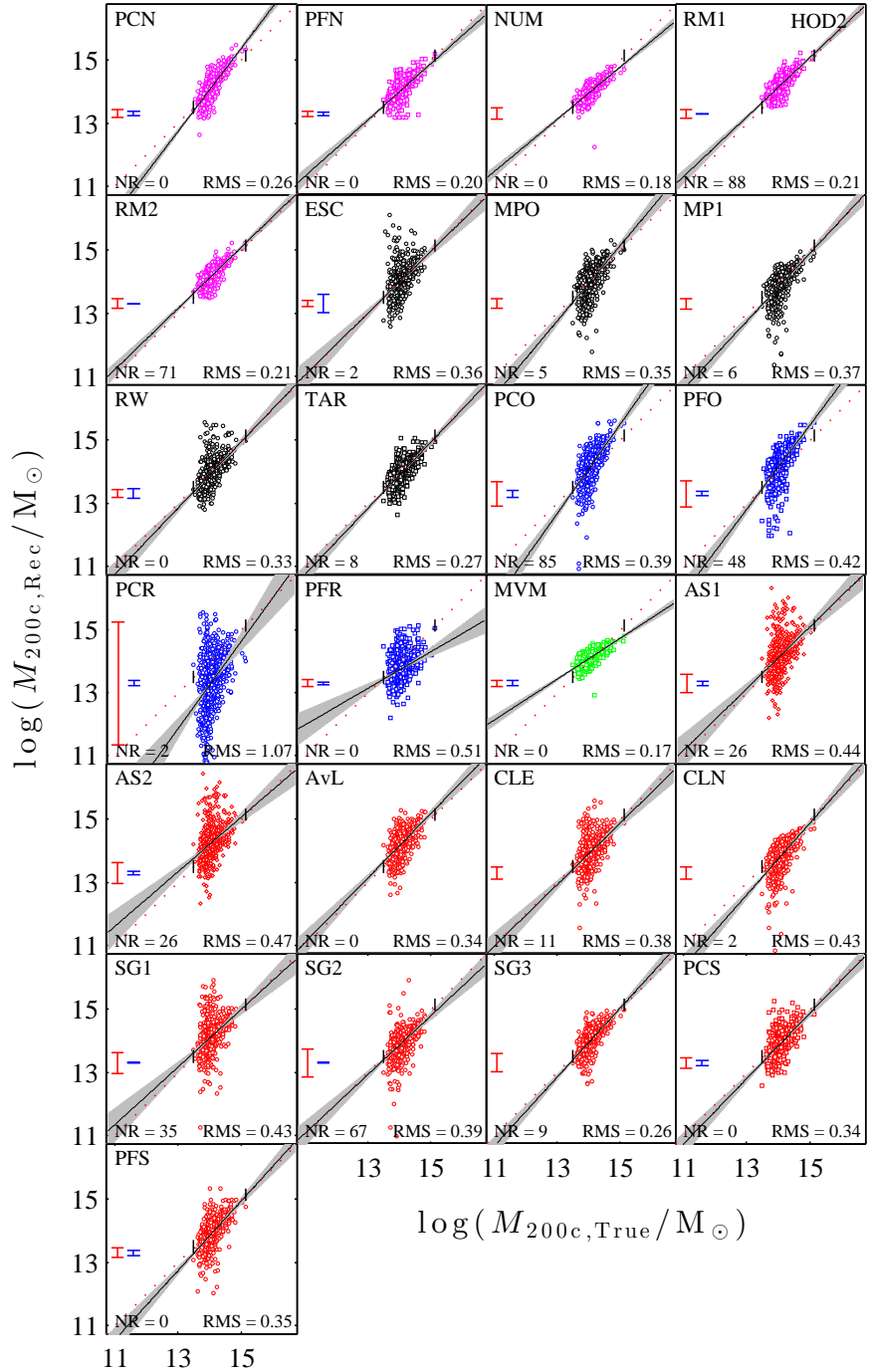}
 \caption{Recovered versus true cluster mass for the 25 methods applied to the HOD2 input catalogue. The colour scheme reflects the approach implemented by each method to deliver a cluster mass from a chosen galaxy membership: magenta (richness), black (phase space), blue (radial), green (abundance-matching) and red (velocity dispersion). The solid black line represents the fit to the recovered log mass produced by the MCMC analysis and the filled grey area presents the $\rm 3\sigma$ boundary of this fit. The red dotted line represents
 the 1:1 relation. `NR' in the legend represents the number of missing groups/clusters. The black ticks that
 lie across the 1:1 relation represent the minimum and maximum
 `true' halo $\log M_{\rm 200c}$. The vertical red bar (left) represents the mean statistical error and the vertical blue bar (right) represents the mean systematic error delivered by methods directly.}
\label{fig:HOD2_SW_Nsigma_outliers_mass_scatter_combined}
\end{figure*}
\clearpage
Each component of this likelihood is weighted by the probability that any given point belongs to either of these distributions:
\begin{eqnarray} 
%\begin{split}
\mathcal{L} &\!\!\!\!=\!\!\!\!& \prod_{i=1,N} p_i \nonumber \\
p_i &\!\!\!\!=\!\!\!\!& \left[(1-P_{\rm b}) P(\log M_{{\rm Rec},i} |  \log
M_{{\rm True},i}
,\sigma_{\log M_{{\rm Rec}, i}},m,c) \right.
\nonumber \\
&\mbox{}& \qquad 
+ \left.
P_{\rm b} P(\log M_{{\rm Rec}, i}|\log M_{{\rm True},i},\sigma_{\rm
  outlier},m,c) \right] \ .
%\end{split}
\end{eqnarray}
Here $P_{\rm b}$ is the posterior fraction of objects belonging to the `bad'
outlier distribution, $\sigma_{M_{\rm Rec, i}}$ is the variance of the `good'
distribution and $m$, $c$ are the slope and intercept of the fit
respectively, which together give the bias at any true or recovered mass. The variance of the `bad' outlier distribution is fixed to be a very large number with the prior that the variance of the `good' distribution
must always be smaller than variance of the `bad' distribution. Flat priors
are adopted for the variance of the `good' distribution, the slope, the
intercept,
while the probability that $N$ data points belong to a `bad' outlier
distribution must be between zero and one.

To efficiently sample our parameter space, we utilise Markov Chain Monte
Carlo (MCMC) techniques that produce posterior probability distributions for
these parameters. In particular, we use the parallel-tempered MCMC sampler
{\sc emcee} \citep{2013PASP..125..306F}. This sampler uses several
ensembles of \emph{walkers} at different \emph{temperatures} to explore the
parameter space. A walker represents a point in the parameter space and at
each iteration of the MCMC, the walkers explore by taking a randomly-sized
step towards another (randomly chosen) walker i.e., towards another point in
parameter space.

Each ensemble of walkers works at a certain `temperature' where the
likelihood is modified, enabling walkers to easily explore different local
maxima i.e., preventing walkers becoming stuck at regions of local instead of
global maxima in the case of a multi-modal likelihood. We employ 50 walkers
at 5 temperatures and perform 2800 iterations, including a `burn-in' of 1500
iterations that are discarded. In total, $50\times5\times2800 = 700\;000$
points in parameter space are sampled for each method and input catalogue. We
use the autocorrelation length, which is a measure of the number of
evaluations of the posterior required to produce independent samples to
verify that convergence has been reached. An example of the marginalised probability
distributions of the parameters produced by this parallel-tempered MCMC can
be found in Figure~\ref{fig:SAM_PCN_MCMC}. Figures of the marginalised probability
distributions of parameters for all methods are available upon request.

Employing various statistics allows us to examine different aspects of the performance of the mass reconstruction methods. The RMS encompasses both scatter and bias and, hence, delivers the overall uncertainty we can expect for our ensemble of mock clusters.  The scatter in the recovered mass, $\sigma_{M_{\rm Rec}}$, delivers a measure of the intrinsic scatter, i.e., in the case of no bias (a slope of unity) and no normalisation/offset. The scatter about the true mass, $\sigma_{M_{\rm True}}$, provides a measure of how well a method performs assuming there is no normalisation/offset in the relationship between recovered and true log mass. Both the scatter in the recovered mass and scatter about the true mass are useful quantities to measure when comparing methods, assuming one could accurately calibrate both the bias and normalisation/offset. The bias at the pivot mass is also calculated, where the pivot mass is taken as the median log mass of the input groups/clusters  sample (log $M_{\rm 200c, true}=14.05$).

%----------------------------------------------------------------
\section{Results and Discussion} 
Now that we have described our analysis procedure, we move on to present the results of the cluster mass estimation comparison. We consider an effective method to be one that minimises scatter, has no bias in the amplitude and the slope of the relation between recovered to true mass and minimises catastrophic outliers and missing groups/clusters (whose masses cannot be determined). We therefore examine several aspects of method performance in the following subsections.

\subsection{Scatter in group/cluster mass recovery} 
Figure~\ref{fig:HOD2_SW_Nsigma_outliers_mass_scatter_combined} shows the recovered versus input log mass for the case of the HOD2 model. The colour scheme reflects the approach implemented by each method to deliver a cluster mass from a chosen galaxy membership, as introduced in Section~\ref{sec:Mass Reconstruction Methods}. These colours are magenta (richness), black (phase space), blue (radial), green (abundance-matching) and red (velocity dispersion). Methods which select an initial cluster membership via the FOF linking method have square shaped markers, phase-space based methods have circle shaped markers and Red Sequence-based methods have diamond shaped markers.

Figure~\ref{fig:HOD2_SW_Nsigma_outliers_mass_scatter_combined} clearly shows
that most methods produce significant scatter for this HOD2 mock galaxy
catalogue. In the case of radial based methods, we find an RMS of at least
0.39~dex up to 1.10~dex, which translates to a factor of 2.5 and 12.6
respectively. We see a better performance of phase-space and velocity
dispersion -based methods -eps-converted-to.pdffor the HOD2 model where scatter is in the range of
0.26~dex up to 0.47~dex, a factor of $\sim$1.8 to 3.0. 
\begin{figure}
 \centering
 \includegraphics[trim = 0mm 0mm 0mm 0mm, clip, width=0.45\textwidth]{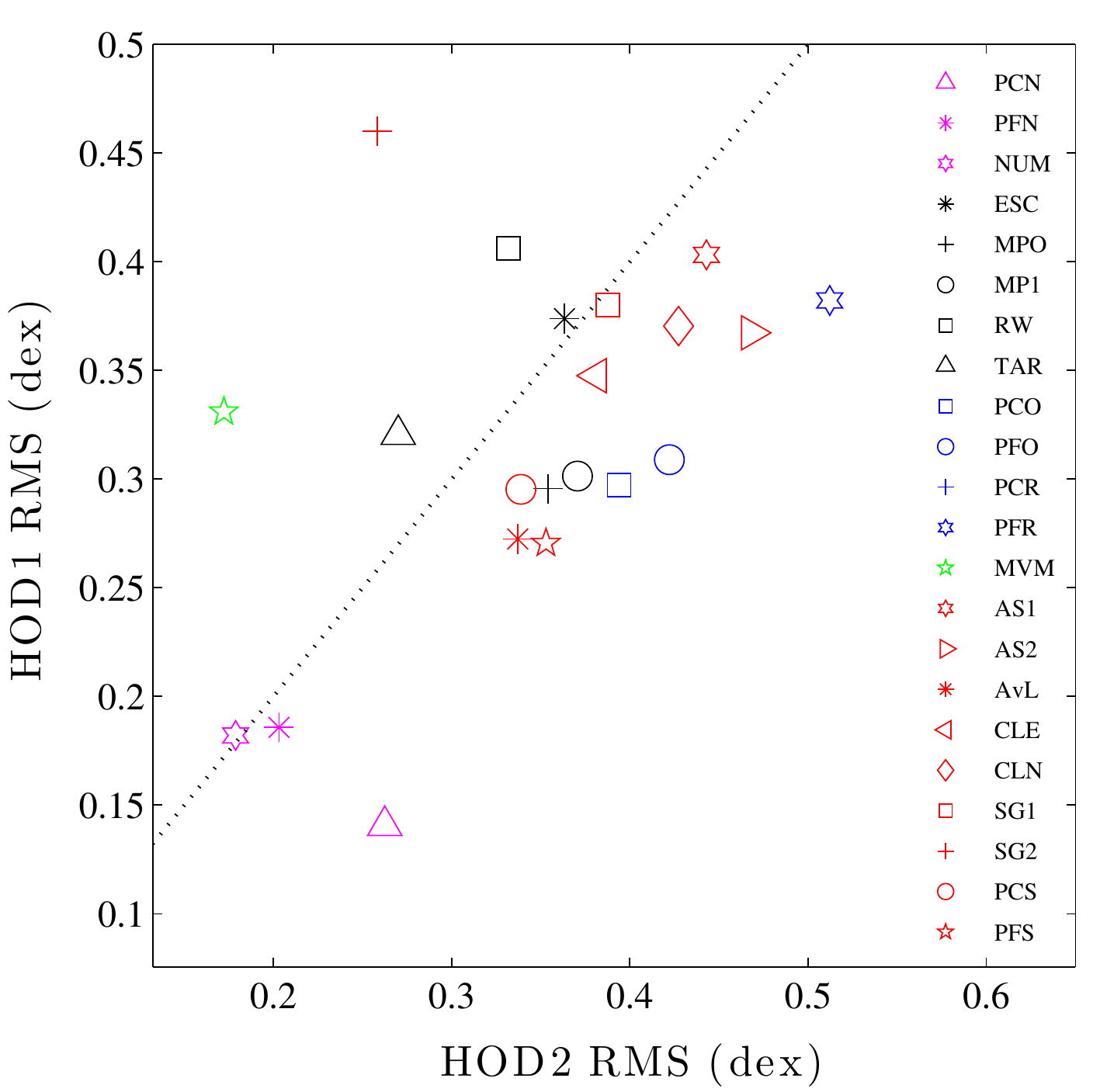}
 \caption{RMS cluster mass accuracies with the simple HOD1 input catalogue versus those derived from the more sophisticated HOD2 catalogue. The dotted black line represents a 1:1 relation. Note: PCR lies beyond the axes of this figure with an RMS of 1.07~dex for HOD2 and 0.74~dex for HOD1.}
\label{fig:HOD1_RMS_vs_HOD2_RMS}
\end{figure}
More traditional richness methods based on simply
counting the number of galaxies out-perform almost all other methods based on
galaxy properties using the HOD2 model.  

\begin{figure*}
 \centering
\includegraphics[trim = 40mm 43mm 50mm 55mm, clip, width=0.83\textwidth]{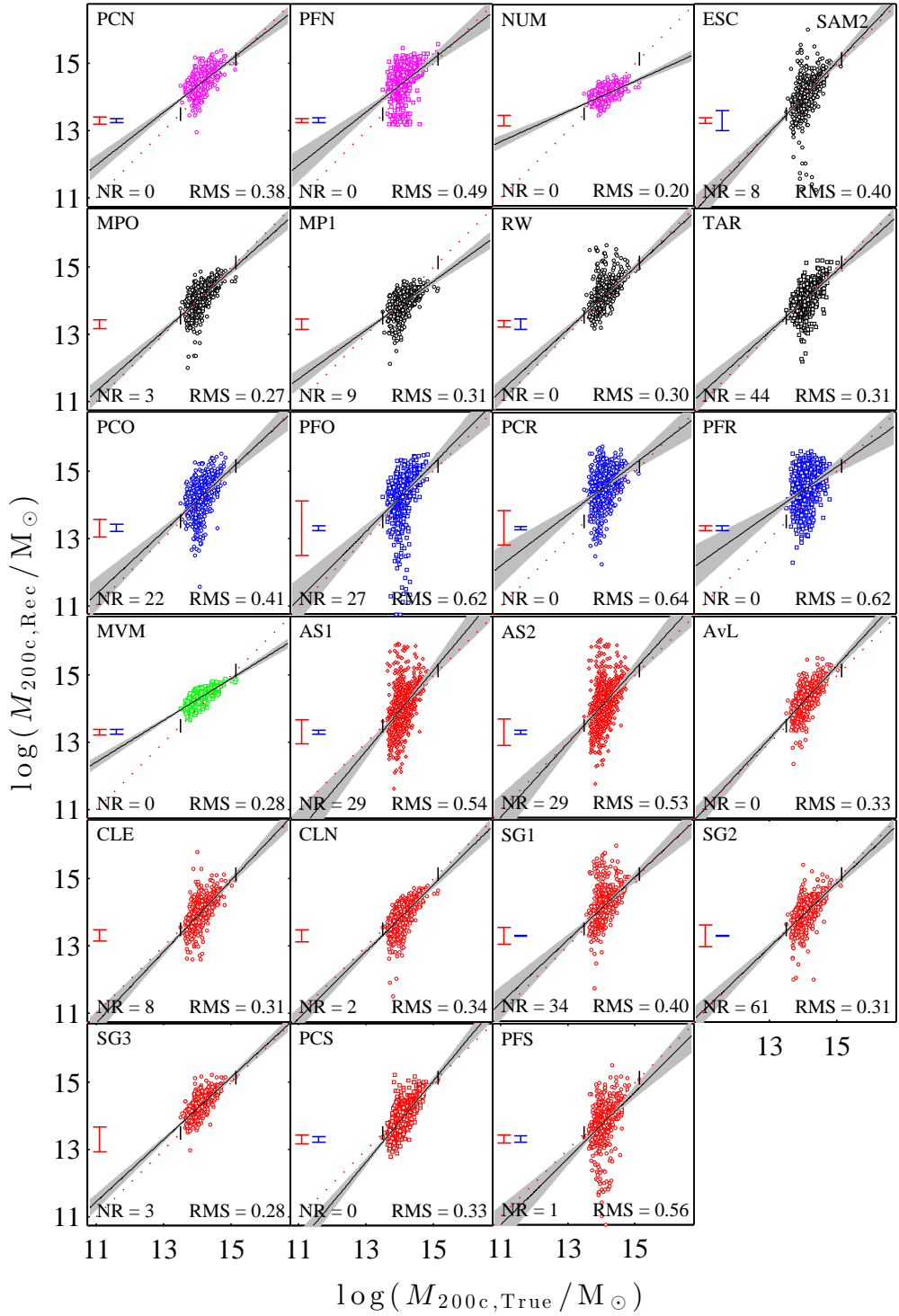}
 \caption{Recovered versus true cluster mass for the 23 methods applied to
   the SAM2 input catalogue. Same 
   notation as in Figure~\ref{fig:HOD2_SW_Nsigma_outliers_mass_scatter_combined}.
  }
\label{fig:SAM2_SW_Nsigma_outliers_mass_scatter_combined}
\end{figure*}

Methods PCN, PFN, NUM, RM1 and RM2
generate much lower scatter of $\sim$0.18 -- 0.26~dex and the abundance
matching based method, MVM, also performs slightly better than the best richness method
(NUM), producing a scatter of 0.17~dex. Note that
according to Poisson statistics and the median number of galaxies in both
catalogues (31), richness-based methods should produce a minimum of scatter
of $\rm 1/(\sqrt{31}\ln 10)=0.08~dex$.
 \begin{figure}
 \centering
 \includegraphics[trim = 0mm 0mm 0mm 8mm, clip, width=0.47\textwidth]{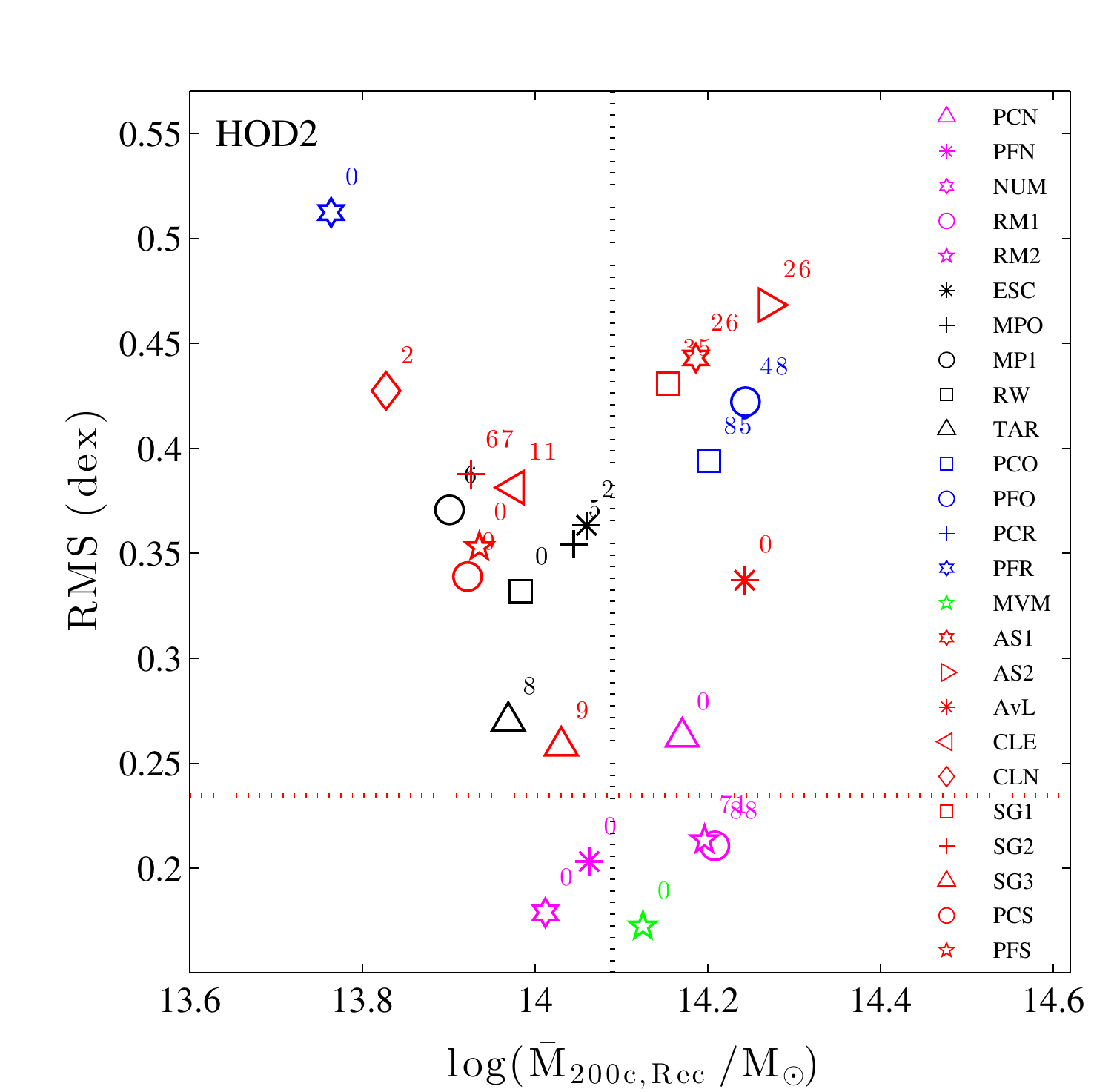}
 \caption{RMS difference in recovered versus true cluster mass versus the mean of
   the recovered log mass distribution, for the 25 methods applied to the HOD2 input catalogue. The dotted black line identifies where the mean of the true mass distribution lies. The red dashed line represents the RMS produced when assuming all clusters have the median true mass of the sample. The number next to each methods' marker represents the number of groups/clusters that are not recovered because they are found to have very low
 ($\rm < 10^{10} \,M_{\odot}$) or zero mass. Note: PCR lies beyond the axes
   of this figure with an RMS of 1.07~dex and $\log \bar{M}_{\rm 200c} =
   13.37$.
}
\label{fig:HOD2_SW_RMS_vs_mean_M200c}
\end{figure}
This can also be compared to the intrinsic scatter of both the HOD2 and SAM2 number of galaxies versus mass, shown in Figure~\ref{fig:HOD2_SAM2_N_vs_M}. The RMS scatter in this relation is 0.09 and 0.12~dex, respectively.
The recovered log mass distributions
for all methods can be seen in
Figure~\ref{fig:HOD2_SW_mass_histograms_combined} along with the true log mass
distributions. Figure~\ref{fig:HOD2_SW_Nsigma_outliers_mass_scatter_combined}
also highlights the importance of the initial galaxy selection stage of mass
estimation. PCR, a method which deduces cluster mass by calculating the RMS
radius of galaxies within a $\rm 1\,Mpc$ aperture and velocity range,
performs poorly without any interloper removal implemented. However, PFR, a
method also based on the RMS radius, is far less affected by the presence of
interloping galaxies as it uses galaxies selected via FOF linking.

As expected, for the majority of methods, the RMS is higher than for Paper I,
as shown Figure~\ref{fig:HOD1_RMS_vs_HOD2_RMS}, where a catalogue based on a
simple HOD model was used (HOD1). Interestingly, there are some methods (abundance matching method MVM,
shifting gapper method SG2, and phase space-based methods RW and TAR) that
actually have lower RMS values for the more complex HOD2 model. When
we examine the residual recovered mass versus true mass for the HOD2
catalogue, shown in Figure~\ref{fig:HOD2_SW_residual_mass_scatter_combined},
it becomes evident that the scatter is substantially higher at lower true
masses, although this effect appears less severe for richness and abundance
matching based methods.
\begin{figure}
 \centering
 \includegraphics[trim = 0mm 0mm 0mm 0mm, clip, width=0.45\textwidth]{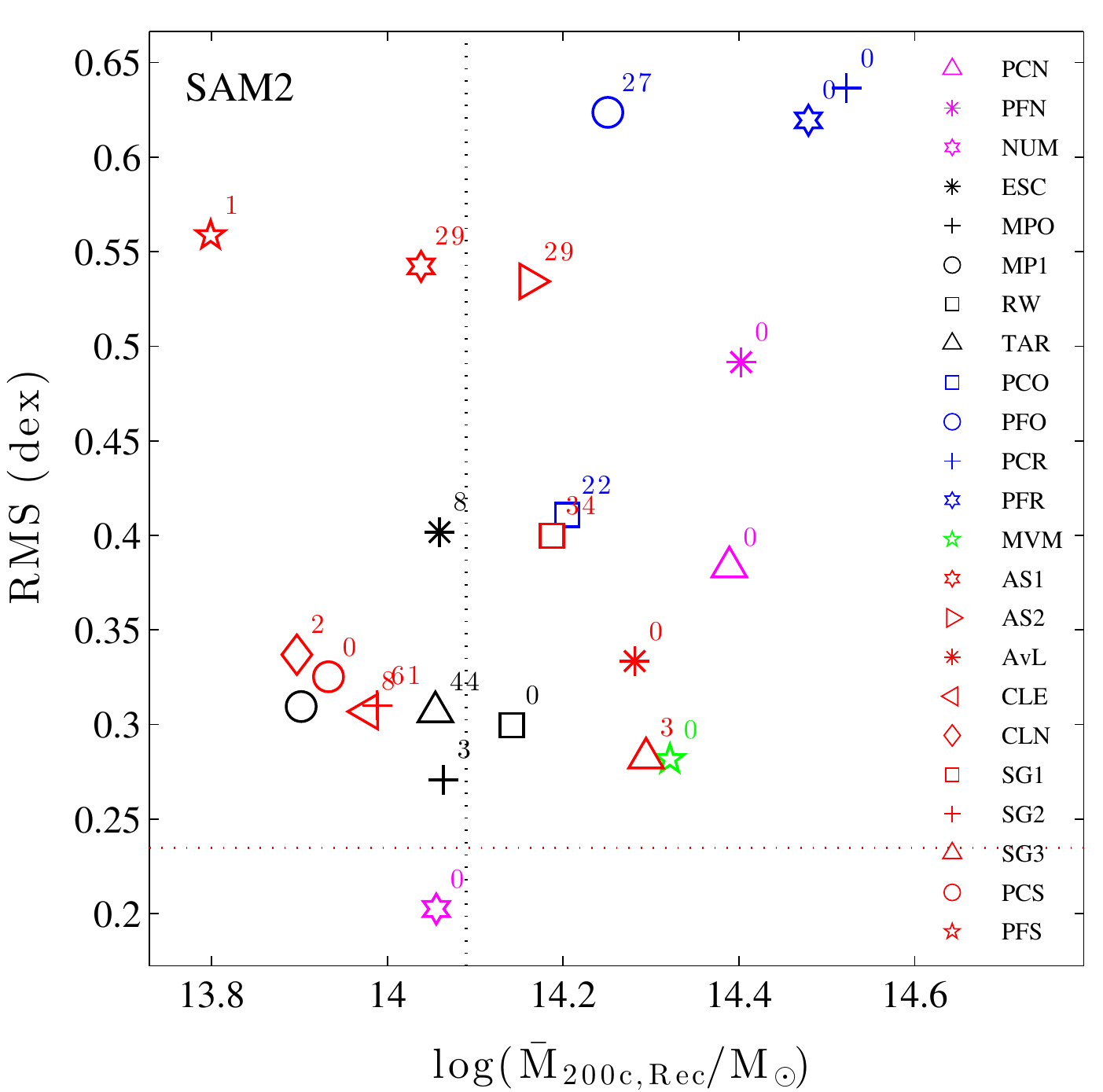}
 \caption{RMS difference in recovered versus true cluster mass versus the mean of
   the recovered log mass distribution, for the 23 methods applied to the SAM2 input catalogue. 
The dotted black line identifies where the mean of the true mass distribution lies. The red dashed line represents the RMS produced when assuming all clusters have the median true mass of the sample. The number next to each methods' marker represents the number of groups/clusters that are not recovered because they are found to have very low
 ($\rm < 10^{10} \,M_{\odot}$) or zero mass.}
\label{fig:SAM2_SW_RMS_vs_mean_M200c}
 \end{figure}
 In addition to the RMS, the scatter in the recovered mass, $\rm
\sigma_{M_{\rm Rec}}$, the scatter about the true mass, $\sigma_{M_{\rm
    True}}$, the slope and the bias at the pivot mass can be seen in
Table~\ref{table:HOD2_SAM2_allmasses_table} for both the HOD2 and SAM2
models. The final column of the two sub-tables shows the merit, a form of
ranking based on the RMS.
\begin{figure}
 \centering
 \includegraphics[trim = 0mm 0mm 0mm 0mm, clip, width=0.44\textwidth]{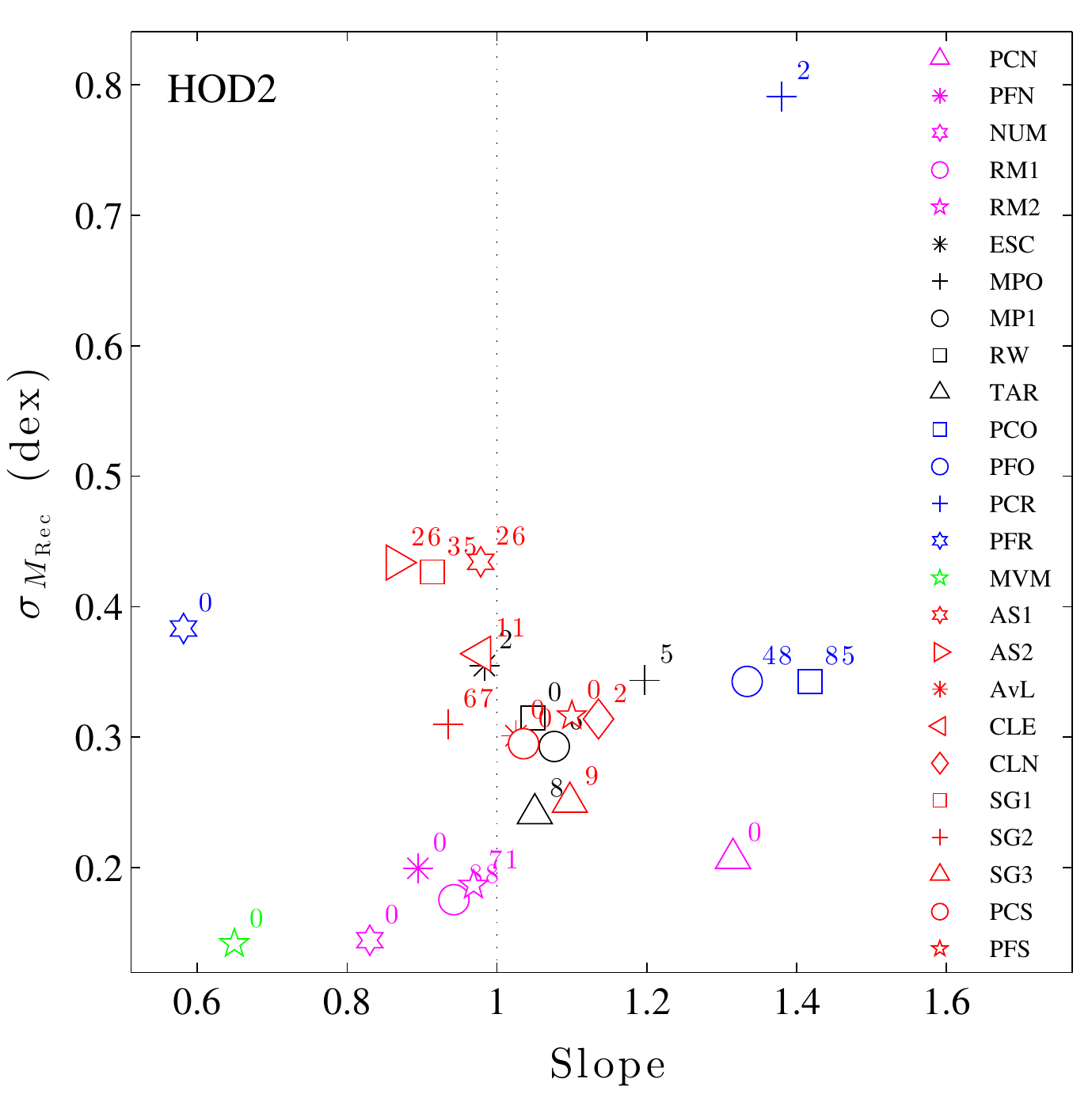}
 \caption{Scatter in the recovered cluster mass versus the slope of the fit to the
   recovered log mass, both delivered by the likelihood analysis, for the 25
   methods applied to the HOD2
   input catalogue. The dotted black line identifies a slope of unity. The number next to each methods' marker represents the number of groups/clusters that are not recovered because they are found to have very low
 ($\rm < 10^{10} \,M_{\odot}$) or zero mass.
}
\label{fig:HOD2_SW_scatter_Mobs_vs_slope}
\end{figure}
\begin{figure}
 \centering
 \includegraphics[trim = 0mm 0mm 0mm 0mm, clip, width=0.45\textwidth]{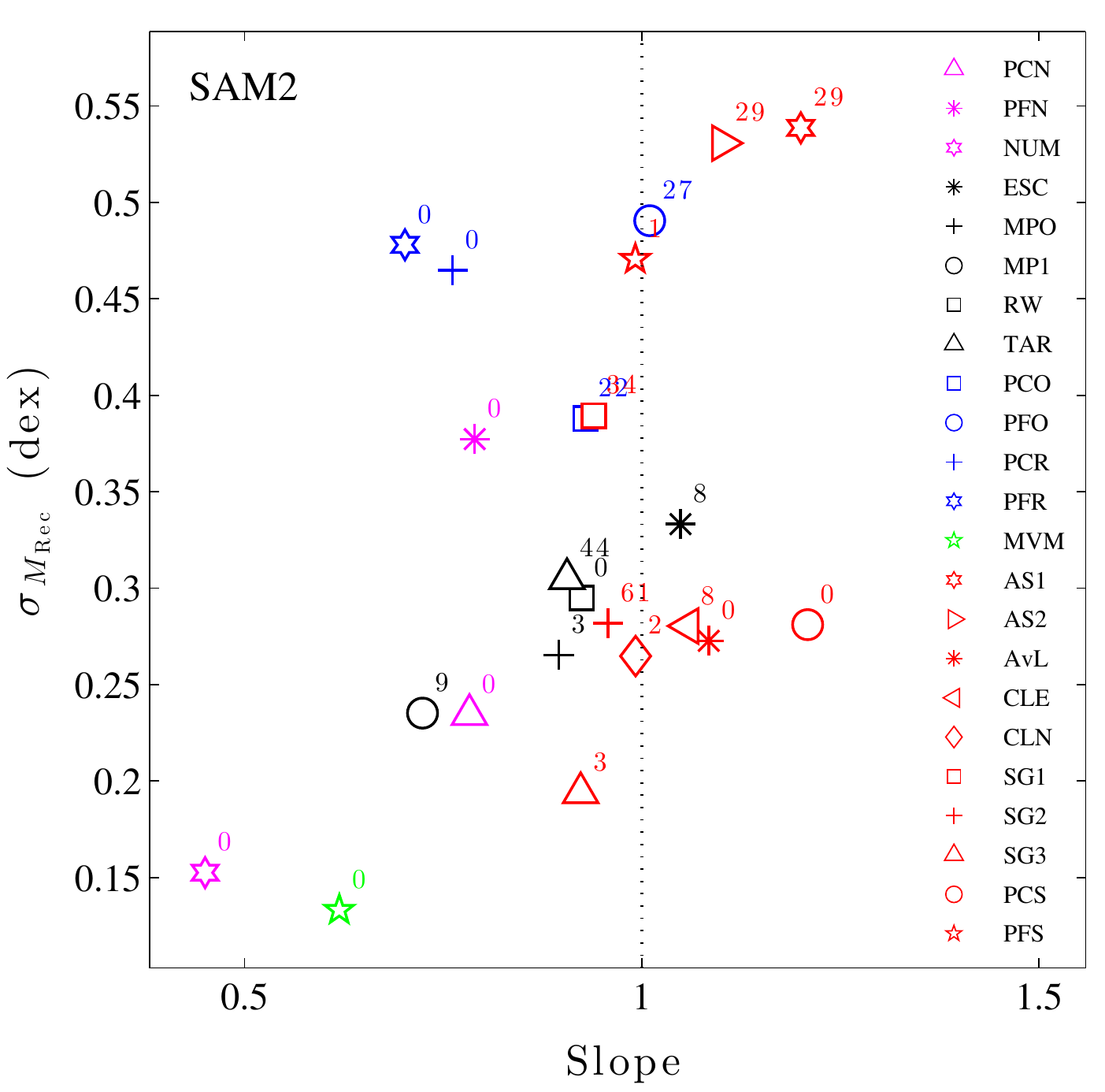}
 \caption{Scatter in the recovered cluster mass versus the slope of the fit to the
   recovered log mass delivered by the likelihood analysis, for the 23
   methods applied to the SAM2 input
   catalogue. This Figure follows the same notation as in Figure~\ref{fig:HOD2_SW_scatter_Mobs_vs_slope}.
}
\label{fig:SAM2_SW_scatter_Mobs_vs_slope}
\end{figure}
It is assigned in different bins: an RMS scatter of
below 0.2~dex is assigned 8 stars and then decreasing numbers of stars are assigned in subsequent
bins of size 0.05~dex. The final bin of methods producing an RMS scatter
greater than 0.5~dex, is given one star. In Paper~I of the project, the
ranking of the methods was not binned but instead, each method was assigned a rank between 1 and 23 corresponding to the lowest and highest RMS. Here, we
assign a merit according to RMS bins to highlight the similarity/disparity
between the scatter produced by different methods in a more linear fashion. 

We
see the same trends in the magnitude of the scatter for different classes of
methods when we look at the scatter in the recovered mass, $\sigma_{M_{\rm
    Rec}}$, and the scatter about the true mass, $\sigma_{M_{\rm
    True}}$. According to these values, if we assume that the 
methods have no bias (slope of unity and zero intercept)  in the relation
between true and recovered mass, then they would deliver scatter as low as
0.14 -- 0.15~dex (MVM and NUM respectively).

\begin{table*} 
\caption{Mass recovery accuracy (RMS, scatter in the recovered mass,
  $\sigma_{M_{\rm Rec}}$, slope, scatter about the true mass $\sigma_{M_{\rm
      True}}$ and the ranking based on $\sigma_{M_{\rm True}}$) and the bias at the pivot mass for all clusters,
for both the HOD2 and SAM2 input galaxy catalogues. The merit is assigned in different bins according to the level of scatter computed by the RMS. A method producing an RMS scatter of below 0.2~dex is assigned 8 stars and then decreasing numbers of stars are assigned in subsequent bins of size 0.05~dex. The final bin of methods producing an RMS scatter greater than 0.5~dex, which is given one star.}
\begin{center}
\begin{tabular}{ l c c c c c l c c c c c l}
\hline
\multicolumn{1}{c}{Method}&
 \multicolumn{6}{c}{HOD2} &
  \multicolumn{6}{c}{SAM2}\\ [0.8ex] 
  \cmidrule(r){2-7} \cmidrule(l){8-13} 
&RMS (dex)&$\sigma_{M_{\rm Rec}}$&Slope&$\sigma_{M_{\rm True}}$& Bias&Merit &RMS (dex)&$\sigma_{M_{\rm Rec}}$&Slope&$\sigma_{M_{\rm True}}$&Bias&Merit\\  [1.0ex] 
\hline
PCN&0.26&0.21&1.32&0.16&$\phantom{-}0.07$&******  &0.38&0.23&0.78&0.30&$\phantom{-}0.31$&****    \\
PFN&0.20&0.20&0.89&0.22&$-0.02$&******* &0.49&0.38&0.79&0.48&$\phantom{-}0.32$&**      \\
NUM&0.18&0.14&0.83&0.17&$-0.07$&********&0.20&0.15&0.45&0.34& $-0.01$&******* \\
RM1&0.21&0.18&0.94&0.19&$\phantom{-}0.12$&******* &&&&&&\\
RM2&0.21&0.19&0.97&0.19&$\phantom{-}0.11$&******* &&&&&&\\
ESC&0.36&0.35&0.98&0.36&$-0.03$&****    &0.40&0.33&1.05&0.32& $-0.01$&***     \\
MPO&0.35&0.34&1.20&0.29&$-0.05$&****    &0.27&0.27&0.90&0.30& $-0.02$&******  \\
MP1&0.37&0.29&1.08&0.27&$-0.19$&****    &0.31&0.24&0.72&0.32& $-0.18$&*****   \\
RW &0.33&0.31&1.05&0.30&$-0.11$&*****   &0.30&0.29&0.92&0.32&$\phantom{-}0.05$&******  \\
TAR&0.27&0.24&1.05&0.23&$-0.12$&******  &0.31&0.31&0.91&0.34& $-0.03$&*****   \\
PCO&0.39&0.34&1.42&0.24&$\phantom{-}0.10$&****    &0.41&0.39&0.93&0.42&$\phantom{-}0.12$&***     \\
PFO&0.42&0.34&1.33&0.26&$\phantom{-}0.15$&***     &0.62&0.49&1.01&0.49&$\phantom{-}0.20$&*       \\
PCR&1.07&0.79&1.38&0.57&$-0.73$&*       &0.64&0.46&0.76&0.61&$\phantom{-}0.44$&*       \\
PFR&0.51&0.38&0.58&0.66&$-0.31$&*       &0.62&0.48&0.70&0.68&$\phantom{-}0.40$&*       \\
MVM&0.17&0.14&0.65&0.22&$\phantom{-}0.05$&********&0.28&0.13&0.62&0.21&$\phantom{-}0.25$&******  \\
AS1&0.44&0.43&0.98&0.44&$\phantom{-}0.10$&***     &0.54&0.54&1.20&0.45& $-0.06$&*       \\
AS2&0.47&0.43&0.87&0.50&$\phantom{-}0.19$&**      &0.53&0.53&1.10&0.48&$\phantom{-}0.07$&*       \\
AvL&0.34&0.30&1.03&0.29&$\phantom{-}0.15$&*****   &0.33&0.27&1.08&0.25&$\phantom{-}0.19$&*****   \\
CLE&0.38&0.36&0.98&0.37&$-0.11$&****    &0.31&0.28&1.06&0.26& $-0.12$&*****   \\
CLN&0.43&0.31&1.14&0.28&$-0.26$&***     &0.34&0.26&0.99&0.27& $-0.19$&*****   \\
SG1&0.43&0.43&0.91&0.47&$\phantom{-}0.07$&***     &0.40&0.39&0.94&0.41&$\phantom{-}0.10$&****    \\
SG2&0.39&0.31&0.94&0.33&$-0.15$&****    &0.31&0.28&0.96&0.29&$-0.10$&*****   \\
SG3&0.26&0.25&1.10&0.23&$-0.06$&******  &0.28&0.19&0.92&0.21&$\phantom{-}0.21$&******  \\
PCS&0.34&0.29&1.04&0.28&$-0.17$&*****   &0.33&0.28&1.21&0.23&$-0.16$&*****   \\
PFS&0.35&0.32&1.10&0.29&$-0.16$&****    &0.56&0.47&0.99&0.47&$-0.29$&*       \\
\hline
\end{tabular}
\end{center}
\label{table:HOD2_SAM2_allmasses_table}
\end{table*}
Now that we have examined the results for the sophisticated HOD2 catalogue, we
move on to examine how well the cluster mass reconstruction methods perform
using the SAM2. Figure~\ref{fig:SAM2_SW_Nsigma_outliers_mass_scatter_combined}
shows the recovered log mass versus input log mass for 23 participating
methods. Note that methods RM1 and RM2 did not participate as the method
could not run on the catalogue due to the less prominent Red Sequence
produced by the SAM2 model. Immediately, we see high levels of scatter for
almost all methods with the exception of NUM. Furthermore, as we saw
with the HOD2 catalogue, this scatter appears significantly worse at lower
group/cluster masses when we look at the residual recovered mass in
Figure~\ref{fig:SAM2_SW_residual_mass_scatter_combined}. Exceptions are for methods
NUM, MP1, TAR and especially PFO, whose scatter is
lower at the low-mass end (though is still comparably large)! Not only does this show that the scatter is
dependent on the true group/cluster mass but it also suggests that this mass
dependence is not consistent across all methods. This is confirmed when we
examine the values of scatter produced when the sample is split into lower
and higher mass groups/clusters, as shown in
Tables~\ref{table:HOD2_splitmasses_table} and
\ref{table:SAM2_splitmasses_table} in the appendix. This implies that if one
had prior knowledge of whether a sample of objects contained either groups or high mass
clusters (e.g., from other cluster mass proxies), then more accurate masses
would be obtained if a method that performed best for that mass category were
chosen, as opposed to a method with lower scatter over the entire mass
range. The recovered log mass distributions for all methods can be seen in
Figure~\ref{fig:SAM2_SW_mass_histograms_combined}  along with
the true log mass distributions.

One can compare the scatter found for MPO and MP1 with the scatter found by
\citet{2013MNRAS.429.3079M} when they tested this method on mock projected
phase space distributions derived from random sampling of the dark matter
particle distribution in haloes of hydrodynamical cosmological
simulations. Indeed, \citet{2013MNRAS.429.3079M} found a scatter of 0.040 and
0.058~dex in log $R_{\rm 200c}$  for samples of 500 and 100 galaxies, which respectively amount to 0.12 and 0.17~dex for log $M_{\rm
  200c}$. In the present study, limited to the higher mass haloes, MPO and
MP1 achieve 0.28 and 0.23~dex scatter with the HOD2 groups/clusters
(Table~\ref{table:HOD2_splitmasses_table}), and 0.22 and 0.21~dex scatter
with SAM2 groups/clusters (Table~\ref{table:SAM2_splitmasses_table}). Given that the present
study uses, on average, lower numbers of galaxies per halo of 38 (HOD2) and
53 (SAM2) for the high-mass subsamples,
 our values appear consistent with the scatter values of
\citet{2013MNRAS.429.3079M}. Similarly, the RW model has been tested by
\citet{2009MNRAS.399..812W}, to yield 0.13~dex for 300 particle haloes, while
in the present study limited to high-mass clusters, RW achieves scatter of
0.27 and 0.26~dex for HOD2 (Table~\ref{table:HOD2_splitmasses_table}) and SAM2
(Table~\ref{table:SAM2_splitmasses_table}), respectively.
 
It is also useful to evaluate the effectiveness of using colours to evaluate the virial masses. For example, the methodology used in MP1 is identical to that of MPO, except that the former is colour-blind. Table~\ref{table:HOD2_SAM2_allmasses_table} indicates that MPO and MP1 have comparable accuracies in mass recovery: for the HOD2 and SAM2 samples, MPO has a scatter 0.05 and 0.01~dex higher than MP1, but an RMS that is 0.03 and 0.09~dex lower. In other words, the colour information helps to reduce the bias, but not the scatter.
%\begin{figure}
% \centering
% \includegraphics[trim = 0mm 0mm 0mm 0mm, clip, width=0.5\textwidth]{SAM2_SW_scatter_Mtrue_vs_slope-eps-converted-to.pdf}
% \caption{SAM2: scatter about the true mass versus slope (bias). The dotted black line identifies a slope of unity. The number next to each methods' marker represents the number of groups/clusters that are not recovered because they are found to have very low
% ($\rm < 10^{10} \,M_{\odot}$) or zero mass.}
%\label{fig:SAM2_SW_scatter_Mtrue_vs_slope}
%\end{figure}

%\begin{figure}
% \centering
% \includegraphics[trim = 0mm 0mm 0mm 0mm, clip, width=0.44\textwidth]{HOD2_slope_vs_SAM2_slope-eps-converted-to.pdf}
% \caption{Slope of the fit to the recovered cluster mass delivered by the
%   likelihood analysis for the SAM2 input catalogue versus that for the HOD2
%   input catalogue. The dotted black line represents a 1:1 relation.
%%\gam{Replace `HOD2' and `SAM2' by `HOD2' and `SAM2'.}
%}
%\label{fig:HOD2_slope_vs_SAM2_slope}
%\end{figure}

\subsection{Bias in group/cluster mass recovery} 
The methods do not collectively under- or over-estimate the
mean true mass for the HOD2 groups/clusters, as shown in
Figure~\ref{fig:HOD2_SW_RMS_vs_mean_M200c}. With the exception of
radial-based techniques, the methods are clustered around the true mean log mass. Interestingly, this agreement was not seen in Paper~I when the methods systematically underestimated the mean log mass when tested
with the more simple HOD1 model. We do not see this underestimation with the
more sophisticated HOD2 model as the calibration of velocity dispersions as a
function of halo mass and radius is updated and the treatment galaxy dynamics
(which affect the spatial and redshift distributions) is slightly
modified. Figure~\ref{fig:HOD2_SW_RMS_vs_mean_M200c} also indicates that
there is no strong correlation between the mean recovered mass and scatter
produced by the methods.

A measure of the bias at the pivot mass, which reflects the bias in the amplitude of the relation between recovered and true log mass, is shown in Table~\ref{table:HOD2_SAM2_allmasses_table}. For the HOD2 catalogue, it is clear that low levels of bias can be produced by many methods: PFN and ESC produce a bias of $\leq \pm 0.03$ whilst other methods MPO, MVM  produce a bias of $\leq \pm 0.05$. Within method classes we see a wide range of these bias values.

For the SAM2 catalogue, we also see that the methods do not collectively under- or over-estimate the mean true mass for the SAM2 groups/clusters, as shown in Figure~\ref{fig:SAM2_SW_RMS_vs_mean_M200c}. We do however, see slightly larger values in the bias at the pivot mass, although methods NUM, ESC, MPO, RW and TAR  produce a bias of $\leq \pm 0.05$. As with the HOD catalogue, we see a wide range of these bias values within method classes. Figures~\ref{fig:HOD2_SW_scatter_Mobs_vs_slope} and 
\ref{fig:SAM2_SW_scatter_Mobs_vs_slope} and the result of a Spearman rank test show that scatter is
 uncorrelated with slope of the recovered and input mass relation for the HOD2 model, however, the scatter is marginally correlated with the slope for the SAM2 (with a p-value = 0.0549). 
%In particular, for the HOD2 catalogue, velocity dispersion based methods
%  produce slopes close to unity. All
%  richness based methods have slopes slightly lower than unity, whereas all but
%  one radial-based methods produce slopes greater than unity. 
Surprisingly, MVM, an
  abundance matching based method with very low scatter in the recovered
  mass, has a low slope of 0.65.
 This flatter slope artificially boosts the scatter about the true mass, as we
 can see from the values in Table~\ref{table:HOD2_SAM2_allmasses_table}. The
 scatter in the recovered mass is 0.14~dex as opposed to 
  for the
 scatter about the true mass. This suggests that if MVM were able to produce a slope equal to
 unity, the scatter in for this method would be as low as 0.14 dex.
% For the SAM2 galaxy catalogues, we see the opposite trend for the radial-based methods, which all produce slopes of $\leq 1$. Richness methods PCN, PFN and NUM also all have slopes substantially lower than unity, while this is not seen for the HOD2 catalogue (note: visually, PFR may appear to have a slope of greater than unity, however, has a slope of less than unity, this is simply due to the density of points in the subplot).

It is important to understand how our results vary due to the underlying
model used to produce the catalogue. As touched on above, we see some
differences in the recovered masses for different classes of methods using
two very different input mock galaxy catalogues. Though we see some
differences method-to-method, collectively, the methods do not systematically
have substantially higher scatter or more bias in the slope or amplitude for
either model. 

\begin{figure*}
 \centering
 \includegraphics[trim = 0mm 0mm 0mm 0mm, clip, width=1.05\textwidth]{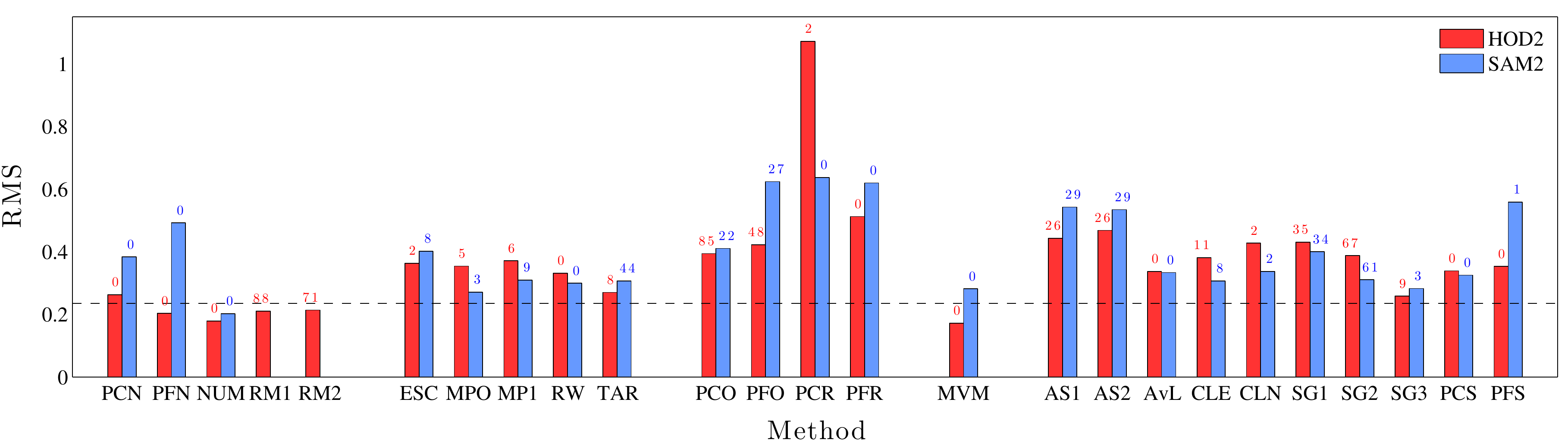}
 \caption{Histogram of the RMS of the
difference between the recovered and true cluster masses (in $\rm
dex$) for all methods applied to the HOD2 (red) and SAM2 (blue) input catalogues. The black dashed line represents the RMS produced when we assume all clusters have the same mass. This uniform mass is chosen to be the mean input log mass. The number above each bar represents the number of missing groups/clusters.}
\label{fig:HOD2_SAM2_M200c_RMS_hist}
\end{figure*}

\begin{figure*}
 \centering
 \includegraphics[trim = 0mm 0mm 0mm 0mm, clip, width=1.05\textwidth]{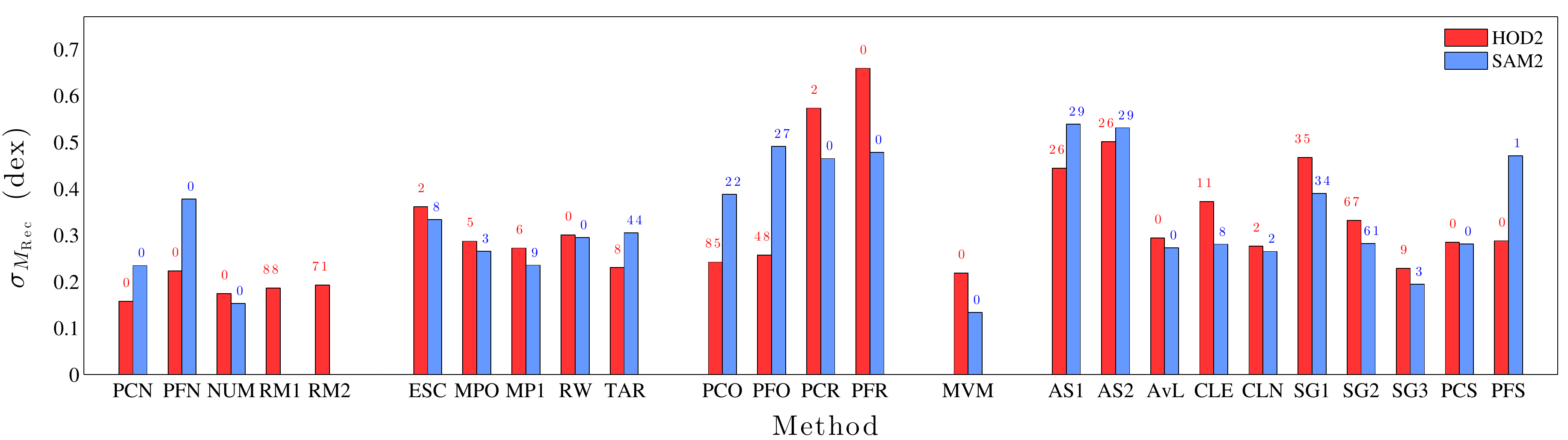}
 \caption{Histogram of the scatter in the recovered cluster mass (in $\rm
dex$) for all methods using the HOD2 (red) and SAM2 (blue) input catalogues. The number above each bar represents the number of missing groups/clusters.}
\label{fig:HOD2_SAM2_M200c_scatter_Mobs_hist}
\end{figure*}

This is especially clear in
Figure~\ref{fig:HOD2_SAM2_M200c_RMS_hist} and
Figure~\ref{fig:HOD2_SAM2_M200c_scatter_Mobs_hist} where histograms of the
RMS and scatter about the true mass are shown for all methods and each
model. There is a surprising similarity between the RMS and scatter in the
recovered mass for both the HOD2 and SAM2 models for many methods. 
This is
encouraging, as it suggests that either the galaxy population produced by
these two contrasting models is analogous or the methods are insensitive to
the differences between these models.

%----------------------------------------------------------------
\subsection{Catastrophic outliers and missing clusters} 

When we examine the performance of these methods, it is clear that there are a number of groups/clusters whose masses are either greatly under- or over-estimated. For example, in the HOD2 catalogue, there are three clusters with mass greater than $\rm 10^{15} \,M_{\odot}$, but some methods predict many more clusters with such large masses (e.g., PCO (39), AS1 (52), AS2 (54) and SG1 (42), see Figure~\ref{fig:SAM2_SW_mass_histograms_combined}). Obtaining the correct number of high mass clusters is crucial for studies selecting high mass clusters -- given the steeply falling  mass function at the high mass end, even a small number of false high mass cluster measurements would have a large impact. 

\begin{figure*}
 \centering
 \includegraphics[trim = 0mm 0mm 0mm 0mm, clip, width=1.05\textwidth]{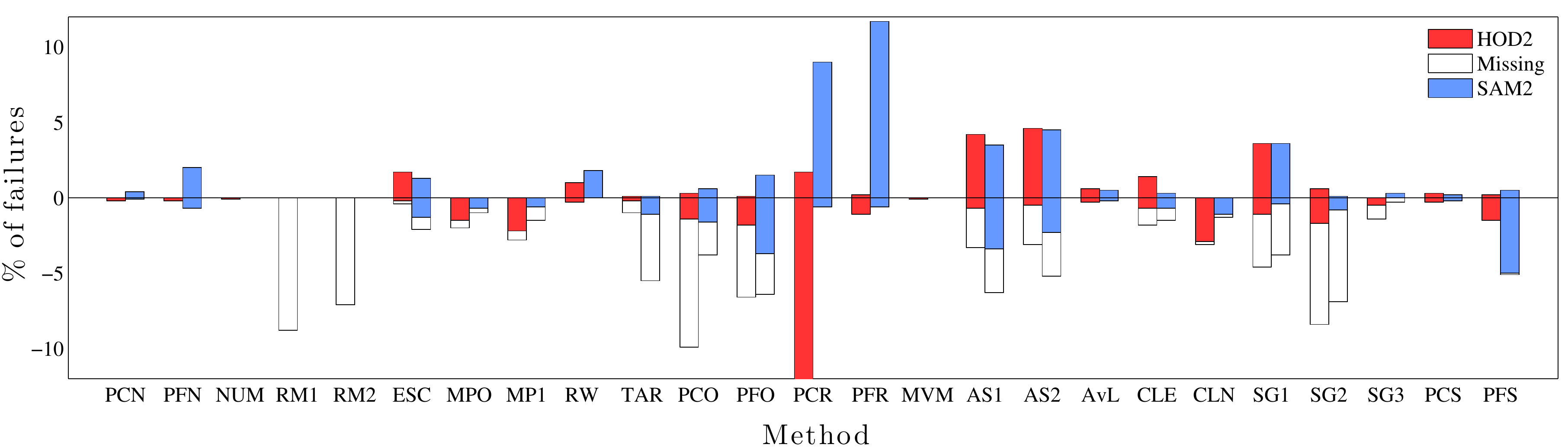}
 \caption{Histogram of the percentage of groups/clusters whose recovered mass
   is over-estimated (positive $\%$) or underestimated (negative $\%$) by a
   factor of ten or more relative to the true mass for all methods applied to
   the HOD2 (red) and SAM2 (blue) input catalogues. The white segments of each bar represents the number of missing groups/clusters. Note that the y-axis of this plot is truncated so that detail at the low percentage range is seen more clearly. PCR falls below this truncation with $32.1\%$ of group/cluster masses underestimated by a factor of ten.}
\label{fig:HOD2_SAM2_M200c_catastrophic_failures}
\end{figure*}

Furthermore, significant underestimations of mass is also very detrimental, as cosmological constraining power increases with lower mass clusters (as signal to noise increases with decreasing mass).  For this reason, it is important to assess the fraction of groups/clusters for which a method under- or over-estimates the mass by a large factor. The percentage of groups/clusters whose mass is under- or over-estimated by a factor of 10 is shown for all methods and each model in the form of histograms in Figure~\ref{fig:HOD2_SAM2_M200c_catastrophic_failures}. Groups/clusters whose masses are over-estimated by over a factor of 10 are shown as a positive percentage and those whose are underestimated by over a factor of 10 are shown as a negative percentage. The percentage of groups/clusters that are missing i.e., no mass was found for these objects, is shown in white. Encouragingly, richness-based methods PCN, NUM, RM1, RM2 and abundance-based method MVM have extremely low fraction of these failures.
\clearpage 
\begin{figure*}
 \centering
\includegraphics[trim = 45mm 52mm 55mm 57mm, clip, width=0.8\textwidth]{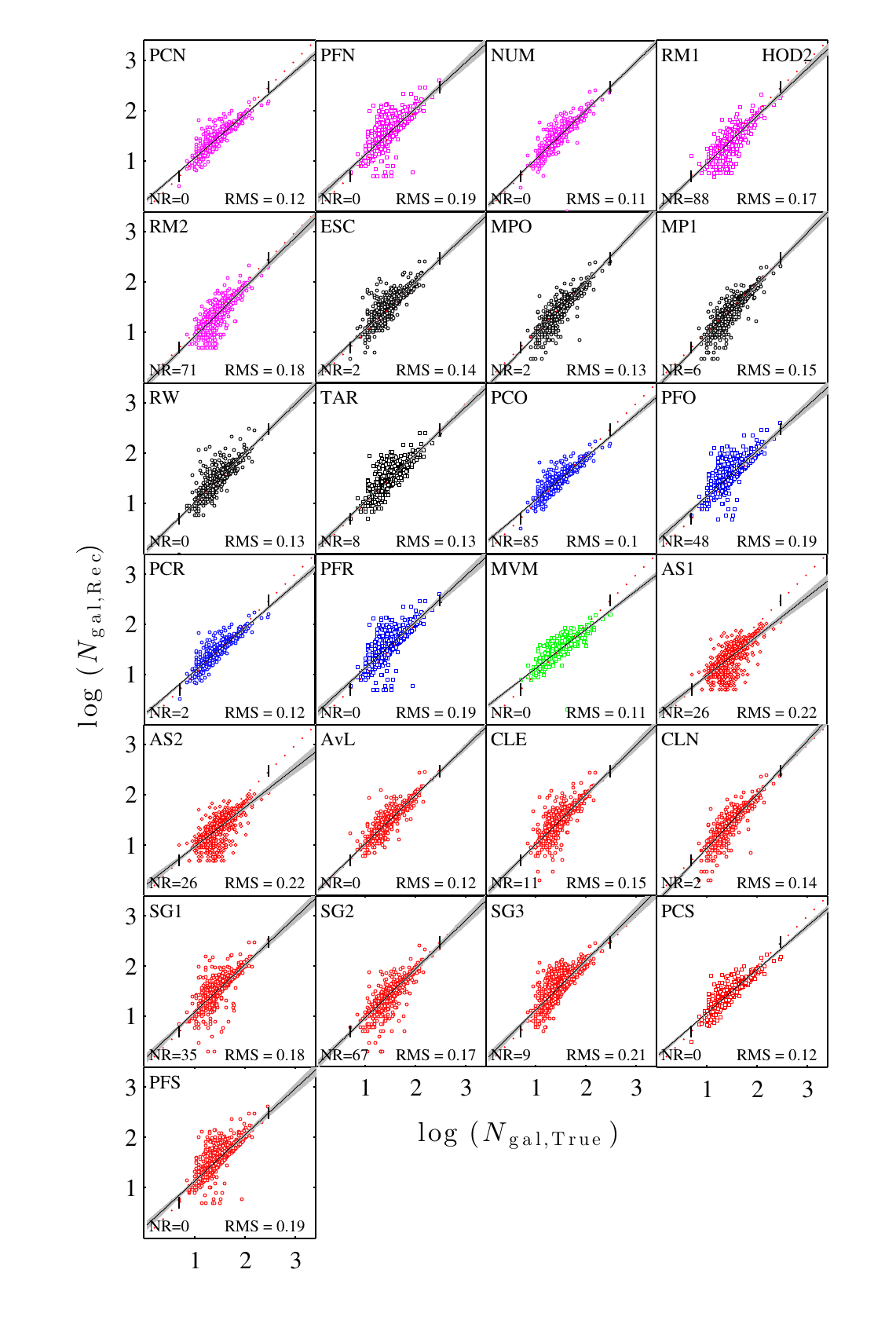}
 \caption{Recovered versus true cluster richness for the 25 methods run on the
   HOD2 input catalogue. The colour scheme reflects the approach implemented by each method to deliver a cluster mass from a chosen galaxy membership: magenta (richness), black (phase space), blue (radial), green (abundance-matching) and red (velocity dispersion). The solid black line represents the fit to the recovered $N_{\rm gal}$ produced by the MCMC analysis and the filled grey area presents the $\rm 3\sigma$ boundary of this fit. The red dotted line represents
 the 1:1 relation. `NR' in the legend represents the number of missing groups/clusters. The black ticks that
 lie across the 1:1 relation represent the minimum and maximum
 `true' halo $N_{\rm gal}$.
}
\label{fig:HOD2_SW_Nsigma_outliers_Ngal_scatter_combined}
\end{figure*}
\begin{figure*}
 \centering
\includegraphics[trim = 42mm 69mm 50mm 55mm, clip, width=0.97\textwidth]{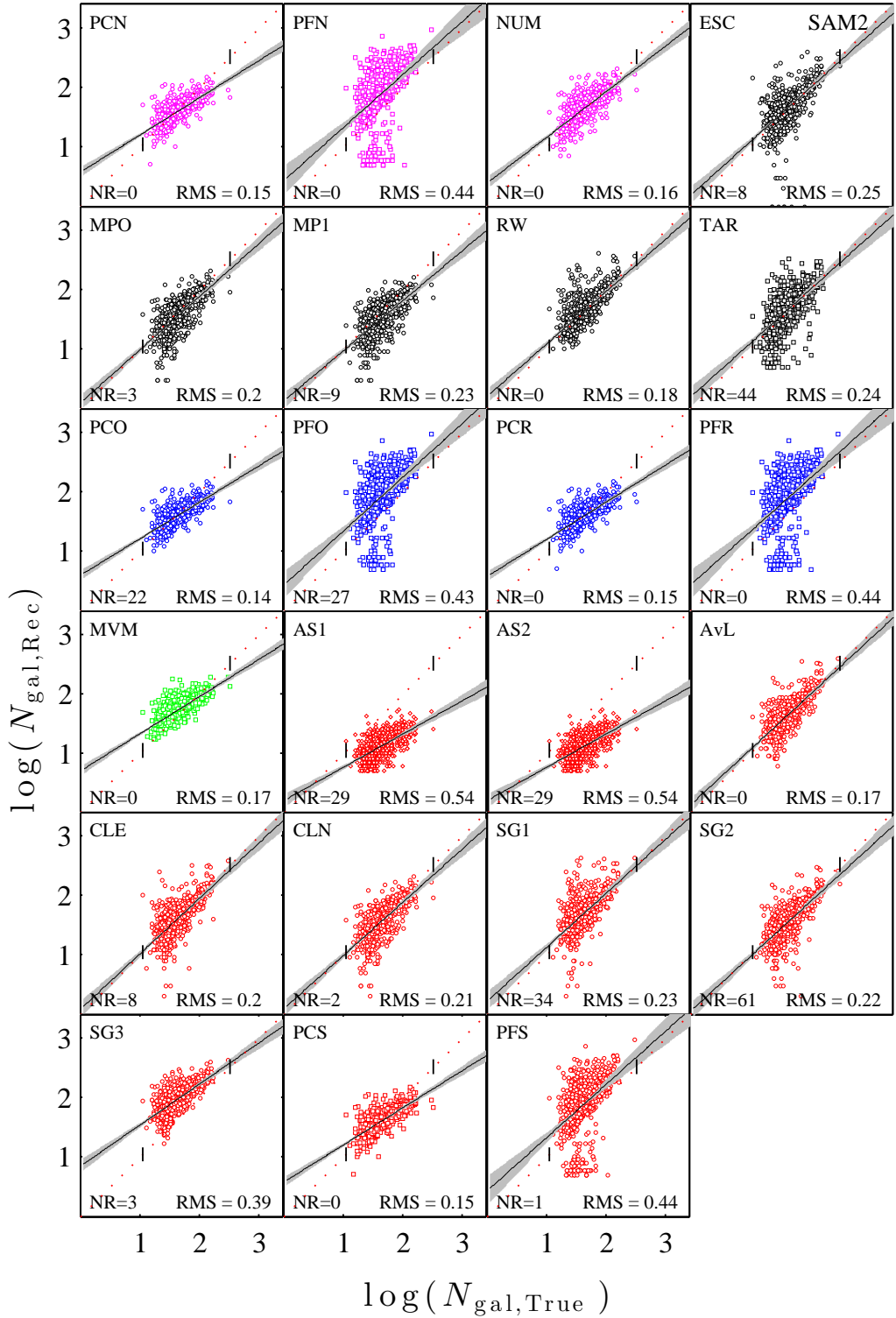}
 \caption{Recovered versus true cluster richness for the 23 methods run on the
   SAM2 input catalogue. This Figure follows the same notation as in
   Figure~\ref{fig:HOD2_SW_Nsigma_outliers_Ngal_scatter_combined}.
}
\label{fig:SAM2_SW_Nsigma_outliers_Ngal_scatter_combined}
\end{figure*}
\clearpage
However, some radial-based and velocity dispersion based methods predict masses out by over a factor of 10 for over $\rm 50$ groups/clusters (i.e., 5\%). This result indicates that if these velocity dispersion, radial-based or phase space -based methods are used, it is vital to also apply abundance matching or certain richness based techniques as a sanity check to ensure there are no catastrophic failures that would misrepresent the shape of the mass function.

The fraction of groups/clusters whose masses are not recovered are shown as white segments in Figure~\ref{fig:HOD2_SAM2_M200c_catastrophic_failures}. We see a large variation between methods, but no correlation of the fraction of missing clusters with method class. Methods such as RM1, RM2, TAR, PCO, PFO and SG2 do not recover masses for $\rm 4 - 8 \%$ of groups/clusters, whereas many methods recover masses for all clusters e.g., PCN, PFN, NUM, PCR, PFR, MVM, AvL, PCS and PFS. 

%----------------------------------------------------------------
\subsection{Group/cluster $N_{\rm gal}$ recovery}
In this section, we present the results of the number of galaxies (i.e., the richness) recovered by the cluster mass reconstruction techniques using both the more sophisticated HOD2 and SAM2 input galaxy catalogues. Figures~\ref{fig:HOD2_SW_Nsigma_outliers_Ngal_scatter_combined} and \ref{fig:SAM2_SW_Nsigma_outliers_Ngal_scatter_combined} show the recovered log number of galaxies versus input log number of galaxies for the case of the HOD2 model and SAM2 model respectively. The colour scheme, lines, symbols and statistics are the same as for the mass comparison figures.
\begin{figure}
 \centering
 \includegraphics[trim = 0mm 0mm 0mm 0mm, clip, width=0.45\textwidth]{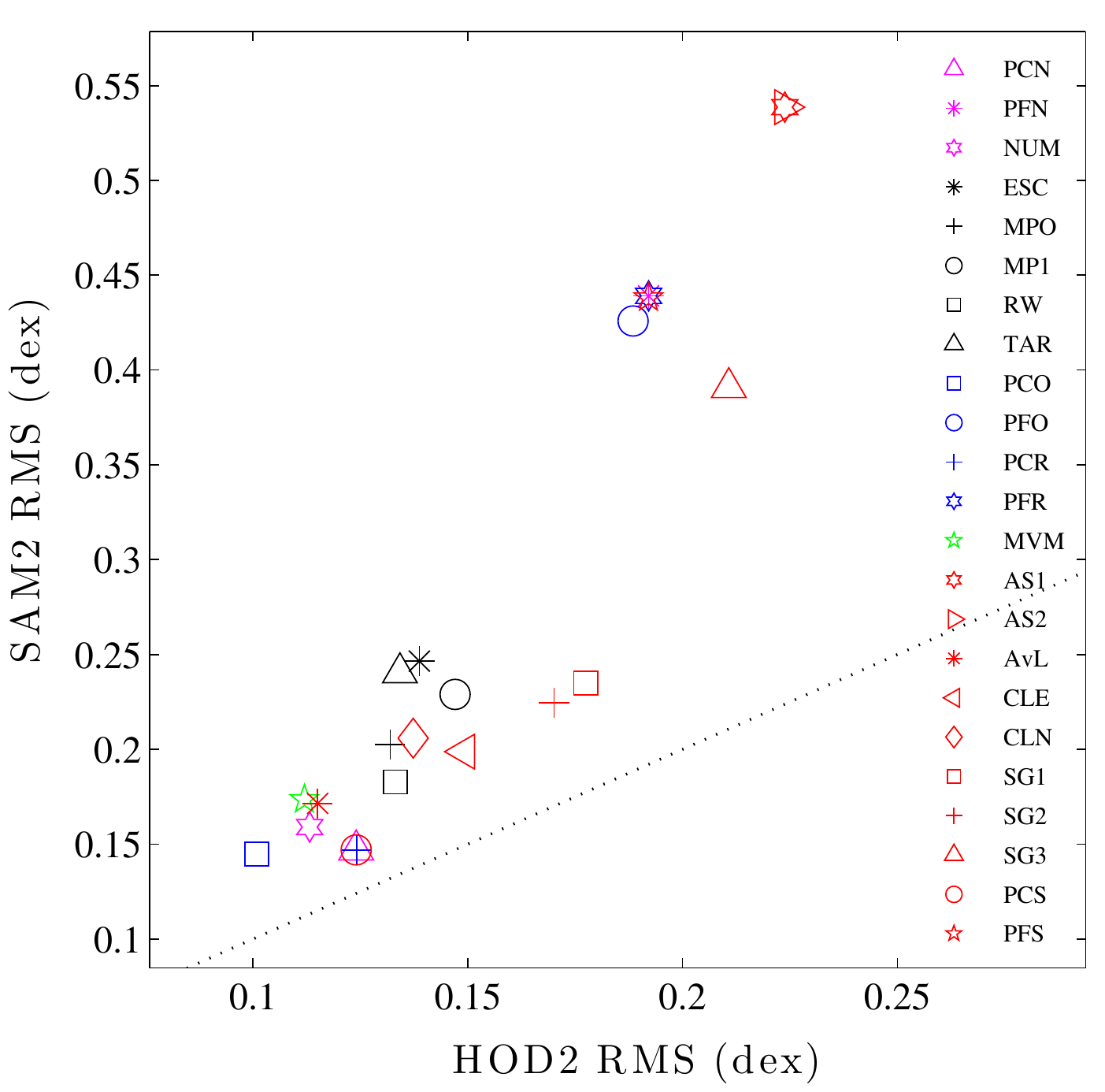}
 \caption{RMS cluster richness errors from the 23 methods applied to the SAM2 input catalogue
   versus those found when applied to the HOD2 input catalogue.
The dotted black line represents a 1:1 relation.
}
\label{fig:HOD2_RMS_vs_SAM2_RMS_Ngal}
\end{figure}

In general, we see a stronger correlation of the recovered richness to the
input richness and lower RMS values for the methods for both the HOD2 and SAM2
catalogues in comparison with group/cluster mass. The mean RMS values produced by
methods using both catalogues are 0.31~dex for mass estimation and 0.21~dex
for $N_{\rm gal}$ estimation respectively. This is also highlighted in
Figure~\ref{fig:HOD2_RMS_vs_SAM2_RMS_Ngal}, which shows the RMS difference between the recovered and input log $N_{\rm gal}$ for the SAM2
catalogue versus the HOD2 catalogue and
Table~\ref{table:HOD2_SAM2_Ngal_table}, which shows this RMS, as well as the scatter in
the recovered richness, $\sigma_{N_{\rm Rec}}$, the scatter about the true
richness, $\sigma_{N_{\rm True}}$, the slope and the bias at the pivot number
of galaxies. 

Again, we see very low RMS values for NUM, MVM and radial-based method PCO for the HOD2 catalogue. The outliers with higher RMS values for the SAM2 catalogue are PFN, PFO,
PFR and PFS, methods that select an initial galaxy membership list via FOF. Red Sequence based methods AS1 and AS2, also have a very high RMS values, though, this is mostly due to the large bias observed at the pivot mass (-0.48~dex). It
is evident from Figure~\ref{fig:HOD2_RMS_vs_SAM2_RMS_Ngal} 
that all methods have lower scatter in the true
number of galaxies for the HOD2 input catalogue in comparison with for the SAM2 input
catalogue. This is not unexpected, as it is
the nature of the HOD2 model to deliver groups/clusters that have a very
strong mass -- richness correlation. Interestingly though, this strong boost
in scatter for the SAM2 catalogue does not necessarily translate to a much
larger scatter in the mass, as reflected in
Figure~\ref{fig:HOD2_SAM2_M200c_RMS_hist}.

Now that we have examined the level of scatter for the recovered richness, we move on to look at the bias. From Table~\ref{table:HOD2_SAM2_Ngal_table}, we see slopes of $\rm < 1$ for the SAM2 catalogue but we do not see the same behaviour for the HOD2 model and, as in the case of the scatter, this does not translate to a systematic shallower slope for the recovered mass. It is important to note that the slope of the $N_{\rm gal}$ --  $M_{\rm 200c}$ relation is lower in the SAM2 as shown in Figure~\ref{fig:HOD2_SAM2_N_vs_M}. We also find that the recovered richness versus recovered mass is as found in Paper~I, where the richness-based methods, have, as expected, very tight relations. In contrast, many other methods have more scatter in both recovered number and recovered mass.
Note that recovering the correct number of galaxies does not necessarily guarantee that the correct member galaxies are being recovered. The fact that we see substantially lower scatter in the recovered number of galaxies but not the mass, indicates that it is not sufficient to simply obtain the correct number of galaxies. To deliver low scatter, it is essential to get the correct membership.
%\begin{figure}
% \centering
% \includegraphics[trim = 0mm 0mm 0mm 0mm, clip, width=0.5\textwidth]{HOD2_slope_vs_SAM2_slope_Ngal-eps-converted-to.pdf}
% \caption{Slope of the fit to the recovered number of galaxies delivered by likelihood analysis for the SAM2 model versus the Phase 2 HOD2 model. The dotted black line represents a 1:1 relation.}
%\label{fig:HOD2_slope_vs_SAM2_slope_Ngal}
%\end{figure}
%----------------------------------------------------------------
\section{Conclusions}
\label{sec: Conclusions}

We have performed an extensive test of 25 different galaxy-based cluster mass
reconstruction methods by using two contrasting mock galaxy catalogues that
are produced using sophisticated, observationally realistic HOD2 and SAM2
models, run on the same halo merger tree extracted from the same cosmological
$N$-body simulation. The aim of this work is to determine the level of
scatter, the bias and completeness that these methods produce, giving insight
into how we can improve on these techniques while generating more realistic
mock galaxy catalogues. The main results are as follows:
\mbox{}
\begin{enumerate}
\item Phase-space and velocity dispersion -based methods deliver a similar
  level of RMS scatter within the range of a factor of $\sim$1.8 -- 3,
  whilst radial-based methods perform significantly worse, delivering an RMS
  scatter of within a factor of $\sim$2.5 -- 12. 
\item Richness
  based methods produce a comparably lower level of RMS scatter within the
  range of a factor of $\sim$1.5 -- 3.1. The lower RMS scatter produced by richness based methods for both HOD2 and SAM2 mock catalogues (where different assumptions are employed to populate dark matter haloes with galaxies) suggest that the good performance of these methods is robust. The abundance matching-based
  technique we tested also produces a comparably lower level of RMS scatter
  within the range of a factor of $\sim$1.5 -- 1.9 for both models.
%\gam{Convert to dex! (You can add ratios in parentheses.)}
\item For many, but not all methods, we find that the scatter is
  group/cluster mass-dependent and the direction of this dependence varies across methods.
\item As expected, for the majority of methods, the scatter is higher than for Paper~I, where 23 methods were tested on a catalogue based on a simple HOD model. Though, interestingly, there are four methods that have lower RMS values for the more complex HOD2 model. 
%\item Methods do not systematically over- or under-estimate cluster masses for either the HOD2 or SAM2 catalogues. This result was not seen in Paper~I where a less sophisticated HOD mock galaxy catalogue was used. This difference is due to the improved treatment of the velocities of the galaxies and highlights the necessity of a more realistic treatment of velocities when generating mock galaxy catalogues. 
\item We see a large variation of bias in the slope of the recovered and input mass relation across all methods for both the HOD2 and SAM2 galaxy catalogues.
\item Many methods produce a significant number of catastrophic failures, where group/cluster masses are over or under -estimated by a factor of $\geq 10$. For studies selecting high mass clusters, these failures can be detrimental due to the steeply falling high-mass end of the cluster mass function. For this reason, we recommend that richness or abundance matching -based methods are used as a sanity check in conjunction with phase-space, velocity dispersion or radial -based methods when high cluster masses are recovered.
\item We see a stronger correlation of the recovered to input number of galaxies for both catalogues in comparison with recovered to input group/cluster mass. The mean RMS produced by methods using both catalogues 0.31~dex for mass estimation and 0.21~dex for $N_{\rm gal}$ estimation. However, this does not mean the correct member galaxies are being selected. The boost in scatter from the number of galaxies to mass indicates that the selection of the correct galaxies (and not just the correct number of galaxies) is a key to delivering lower scatter for these methods.
\item We see a variation of bias in the slope of the recovered and input number of galaxies relation across all methods for the HOD2, however, all methods produce a slope of less than unity for the SAM2 galaxy catalogue.  
\item Though we see some differences method-to-method, in general, methods do not have significantly higher scatter for either the more sophisticated HOD2 or the SAM2 galaxy catalogues. This is encouraging, as it suggests that either the galaxy population produced by these two contrasting models is analogous or the methods are insensitive to the differences between these models.
\end{enumerate}

There are several outstanding questions that we hope to address in future
using our dataset. What is the impact of observational limitations such as
fibre collisions or survey artifacts on group/cluster membership and hence mass
recovery? What is the impact of halo shape and
concentration
%and concentration (Mamon et al. in preparation) 
on group/cluster mass recovery  (Wojtak et al. in preparation)? What produces the catastrophic under- or over- estimates in each of the 25 methods? These projects share the overall goal of improving or constructing more accurate cluster mass reconstruction techniques.

%----------------------------------------------------------------
\section*{Acknowledgments}
We would like to acknowledge funding from the Science and Technology Facilities Council (STFC). DC would like to thank the
Australian Research Council for receipt of a QEII Research
Fellowship. The Dark Cosmology Centre is funded by the Danish National Research Foundation. The authors would like to express special thanks to the Instituto de Fisica Teorica (IFT-UAM/CSIC in Madrid) for its hospitality and support, via the Centro de Excelencia Severo Ochoa Program under Grant No. SEV-2012-0249, during the three week workshop ``nIFTy Cosmology" where this work developed. We further acknowledge the financial support of the University of Western 2014 Australia Research Collaboration Award for ``Fast Approximate Synthetic Universes for the SKA", the ARC Centre of Excellence for All Sky Astrophysics (CAASTRO) grant number CE110001020, and the two ARC Discovery Projects DP130100117 and DP140100198. We also recognise support from the Universidad Autonoma de Madrid (UAM) for the workshop infrastructure. RAS acknowledges support from the NSF grant AST-1055081. CS acknowledges support from the European Research Council under FP7 grant number 279396. SIM acknowledges the support of the STFC consolidated grant (ST/K001000/1) to the astrophysics group at the University of Leicester.
ET acknowledge the support from the ESF grant IUT40-2. The authors contributed in the following ways to this paper: LO, RAS, FRP, \& DC designed and organised this project. LO performed the analysis presented and wrote the majority of the paper. LO, ET, SIM, RP, TP \& FRP organised the workshop that initiated this project. MRM \& SB, contributed to the analysis. The other authors (as listed in section 3) provided results and descriptions of their respective algorithms.
\bibliographystyle{mn2e}
\bibliography{HMRC_phase2_bibliography}

%----------------------------------------------------------------
%\newpage
%\appendix
%\onecolumn
\newpage
\begin{table*}

\appendix
\begin{flushleft}
\section{Properties of the Mass Reconstruction Methods}
\end{flushleft}
 \centering
 \caption{Illustration of the member galaxy selection process for all methods. The second column details how each method selects an initial member galaxy sample, while the third column outlines the member galaxy sample refining process. Finally, the fourth column describes how methods treat interloping galaxies that are not associated with the clusters.}
 \begin{tabular}{p{0.25cm} p{4.0cm} p{5cm} p{4.5cm}}
 \toprule
 %% \multirow{2}[3]{*}{\textbf{Methods}}&\multicolumn{3}{c}{Member galaxy
 %%   selection methodology} \\[1.0ex]
 \multirow{2}{1cm}{\textbf{Methods}}&\multicolumn{3}{c}{Member galaxy selection methodology} \\[1.0ex]
 \cline{2-4}
 &Initial Galaxy Selection&Refine Membership&Treatment of Interlopers \\
 \midrule
 \textcolor{magenta}{\textbf{PCN}}&Within $\rm 5\,Mpc$, $\rm 1000\,km\,s^{-1}$&Clipping of $\pm3\,\sigma$, using galaxies within $\rm 1\,Mpc$&Use galaxies at $\rm 3-5 \,Mpc$ to find interloper population to remove \\
 \textcolor{magenta}{\textbf{PFN}}&FOF&No&No \\
 
 \textcolor{magenta}{\textbf{NUM}}&Within $\rm 3\,Mpc$, $\rm 4000\,km\,s^{-1}$
&1) Estimate $R_{\rm 200c}$  from the relationship between $R_{\rm 200c}$  and
 richness deduced from CLE; 2) Select galaxies within $R_{\rm 200c}$  and with
  $|v|<2.7\,\sigma_{\rm los}^{\rm NFW}(R)$&No \\
  \textcolor{magenta}{\textbf{RM1}}&Red Sequence&Red Sequence& Probabilistic \\
  \textcolor{magenta}{\textbf{RM2}}&Red Sequence&Red Sequence& Probabilistic \\
\
 \textcolor{black}{\textbf{ESC}}&Within preliminary $R_{\rm 200c}$  estimate and $\rm \pm3500 \,km\,s^{-1}$&Gapper technique&Removed by Gapper technique\\
 
 \textcolor{black}{\textbf{MPO}}&Input from CLN&1) Calculate $R_{\rm 200c}$ , $R_{\rm \rho}$, $R_{\rm red}$, $R_{\rm blue}$ by MAMPOSSt method; 2) Select members within radius according to colour&No \\
 
 \textcolor{black}{\textbf{MP1}}&Input from CLN&Same as MPO except colour blind&No \\
 
 \textcolor{black}{\textbf{RW}}&Within $\rm 3\,\,Mpc$, $\rm 4000\,\,km\,s^{-1}$&Within $R_{\rm 200c}$ , $|2\Phi(R)|^{1/2}$, where $R_{\rm 200c}$  obtained iteratively
&No \\

 \textcolor{black}{\textbf{TAR}}&FOF&No&No \\
 
 \textcolor{blue}{\textbf{PCO}}&Input from PCN& Input from PCN&Include interloper contamination in density fitting \\
 
 \textcolor{blue}{\textbf{PFO}}&Input from PFN&Input from PFN&No \\
 
 \textcolor{blue}{\textbf{PCR}}&Input from PCN&Input from PCN&Same as PCN \\
 
 \textcolor{blue}{\textbf{PFR}}&Input from PFN&Input from PFN&No \\

 \textcolor{green}{\textbf{MVM}}&FOF (ellipsoidal search range, centre
of most luminous galaxy)&Increasing mass limits, then FOF, loops until
closure condition&No \\
 
 \textcolor{red}{\textbf{AS1}}&Within $\rm 1\,Mpc$, $\rm 4000\,km\,s^{-1}$, constrained by colour-magnitude relation&Clipping of $\pm3\,\sigma$&Removed by clipping of $\pm3\,\sigma$ \\
 
 \textcolor{red}{\textbf{AS2}}&Within $\rm 1\,Mpc$, $4\rm 000\,km\,s^{-1}$, constrained by colour-magnitude relation&Clipping of $\pm3\,\sigma$&Removed by clipping of $\pm3\,\sigma$ \\
 
 \textcolor{red}{\textbf{AvL}}&Within $2.5\,\sigma_{v}$ and $0.8\,R_{\rm 200}$&Obtain $R_{\rm 200c}$  and $\,\sigma_{v}$ by $\,\sigma$-clipping&No \\
 
 \textcolor{red}{\textbf{CLE}}&Within $\rm 3\,Mpc$, $\rm 4000\,km\,s^{-1}$&1)
 Estimate $R_{\rm 200c}$  from the aperture velocity dispersion; 2) Select galaxies within
 $R_{\rm 200c}$  and with $|v|<2.7\,\sigma_{\rm los}^{\rm NFW}(R)$; 3) Iterate steps 1 and 2 until convergence&Obvious interlopers are removed by velocity gap technique, then further treated in iteration by $\sigma$ clipping \\
 
 \textcolor{red}{\textbf{CLN}}&Input from NUM&Same as CLE&Same as CLE \\
 
 \textcolor{red}{\textbf{SG1}}&Within $\rm 4000 \,km\,s^{-1}$&1) Measure $\,\sigma_{\rm gal}$, estimate $M_{\rm 200c}$ and $R_{\rm 200c}$ ; 2) Select galaxies within $R_{\rm 200c}$ ; 3) Iterate steps 1 and 2 until convergence&Shifting gapper with minimum bin size of $\rm 250 \,kpc$ and 15 galaxies; velocity limit $\rm 1000\,km\,s^{-1}$ from main body \\
 
 \textcolor{red}{\textbf{SG2}}&Within $\rm 4000 \,km\,s^{-1}$&1) Measure $\sigma_{\rm gal}$, estimate $M_{\rm 200c}$ and $R_{\rm 200c}$ ; 2) Select galaxies within $R_{\rm 200c}$ ; 3) Iterate steps 1 and 2 until convergence&Shifting gapper with minimum bin size of $\rm 150 \,kpc$ and 10 galaxies; velocity limit $\rm 500 \,km\,s^{-1}$ from main body \\
  \textcolor{red}{\textbf{SG3}}&Within $\rm 2.5\,h^{-1}\,Mpc$ and $\rm   4000\,km\,s^{-1}$. Velocity distribution symmetrized& Measure $\,\sigma_{\rm gal}$, correct for velocity errors, then estimate $M_{\rm 200c}$ and $R_{\rm 200c}$  and apply the
surface pressure term correction&Shifting gapper with minimum bin size of
$\rm 420\,h^{-1} kpc$ and 15 galaxies\\
 \textcolor{red}{\textbf{PCS}}&Input from PCN&Input from PCN&Same as PCN \\
 \textcolor{red}{\textbf{PFS}}&Input from PFN&Input from PFN&No \\
 \bottomrule
 \end{tabular}
 \label{table:appendix_table_1}
\end{table*}
\begin{table*}
\renewcommand\thetable{A2} 
 \centering
 \centering
 \caption{Characteristics of the mass reconstruction process of methods used in this comparison. The second, third, fourth and fifth columns illustrate whether a method calculates/utilises the velocities, velocity dispersion, radial distance of galaxies from cluster centre, the richness and the projected phase space information of galaxies respectively. If a method assumed a mass or number density profile it is indicated in columns six and seven.}
 \begin{tabular}{c c c c c c c c}
 \toprule
 \multirow{2}[4]{*}{\textbf{Methods}}&\multicolumn{7}{c}{Galaxy properties used to obtain group/cluster membership and estimate mass} \\[1.0ex]
 \cline{2-8}
 &Velocities&Velocity dispersion&Radial distance&Richness&Projected phase space&Mass density profile&Number density profile\\
 \hline
 \textcolor{magenta}{\textbf{PCN}}&Yes&No&No&Yes&No&No&No \\
 \textcolor{magenta}{\textbf{PFN}}&Yes&No&No&Yes&No&No&No\\
 \textcolor{magenta}{\textbf{NUM}}&No&No&No&Yes&Yes&No&No\\
 \textcolor{magenta}{\textbf{RM1}}&No& No& Yes& Yes &No&No& NFW \\
 \textcolor{magenta}{\textbf{RM2}}&No&No& Yes&Yes&No&No& NFW \\
  
 \textcolor{black}{\textbf{ESC}}&Yes&Yes&Yes&No&No&Caustics&No \\
 \textcolor{black}{\textbf{MPO}}&Yes&No&Yes&No&Yes&NFW&Yes\\
 \textcolor{black}{\textbf{MP1}}&Yes&No&Yes&No&Yes&NFW&Yes\\
 \textcolor{black}{\textbf{RW}}&Yes&No&Yes&No&Yes&NFW&Yes\\
 \textcolor{black}{\textbf{TAR}}&Yes&Yes&Yes&No&No& NFW&No\\
 
 \textcolor{blue}{\textbf{PCO}}&Yes&No&No&No&No&NFW&Yes\\
 \textcolor{blue}{\textbf{PFO}}&Yes&No&No&No&No&NFW&Yes\\
 \textcolor{blue}{\textbf{PCR}}&Yes&No&Yes&No&No&No&No\\
 \textcolor{blue}{\textbf{PFR}}&Yes&No&Yes&No&No&No&No \\
  
 \textcolor{green}{\textbf{MVM}}&Yes&Yes&Yes&No&No&NFW&No\\
 
 \textcolor{red}{\textbf{AS1}}&Yes&Yes&No&No&No&No&No\\
 \textcolor{red}{\textbf{AS2}}&Yes&No&Yes&No&Yes&No&No \\
 
 \textcolor{red}{\textbf{AvL}}&Yes&Yes&Yes&No&No&No&No\\
 
 \textcolor{red}{\textbf{CLE}}&Yes&Yes&No&No&No&NFW&NFW\\
 
 \textcolor{red}{\textbf{CLN}}&Yes&Yes&No&No&No&NFW&NFW\\
 
 \textcolor{red}{\textbf{SG1}}&Yes&Yes&Yes&No&No&No&No\\
 
 \textcolor{red}{\textbf{SG2}}&Yes&Yes&Yes&No&No&No&No\\
 
 \textcolor{red}{\textbf{SG3}}&Yes&Yes&Yes&No&No&No&No\\

 \textcolor{red}{\textbf{PCS}}&Yes&Yes&No&No&No&No&No \\
 
 \textcolor{red}{\textbf{PFS}}&Yes&Yes&No&No&No&No&No \\
 \bottomrule \\
 \end{tabular} 
  \label{table:appendix_table_2}
\end{table*}
\vspace{5mm}
\setcounter{figure}{0} \renewcommand{\thefigure}{B\arabic{figure}}
\begin{figure*}
\begin{flushleft}
\section*{Appendix B: Richness and velocity dispersion -- mass relations}
\end{flushleft}
 \centering
 \includegraphics[trim = 0mm 0mm 0mm -7mm, clip, width=0.46\textwidth]{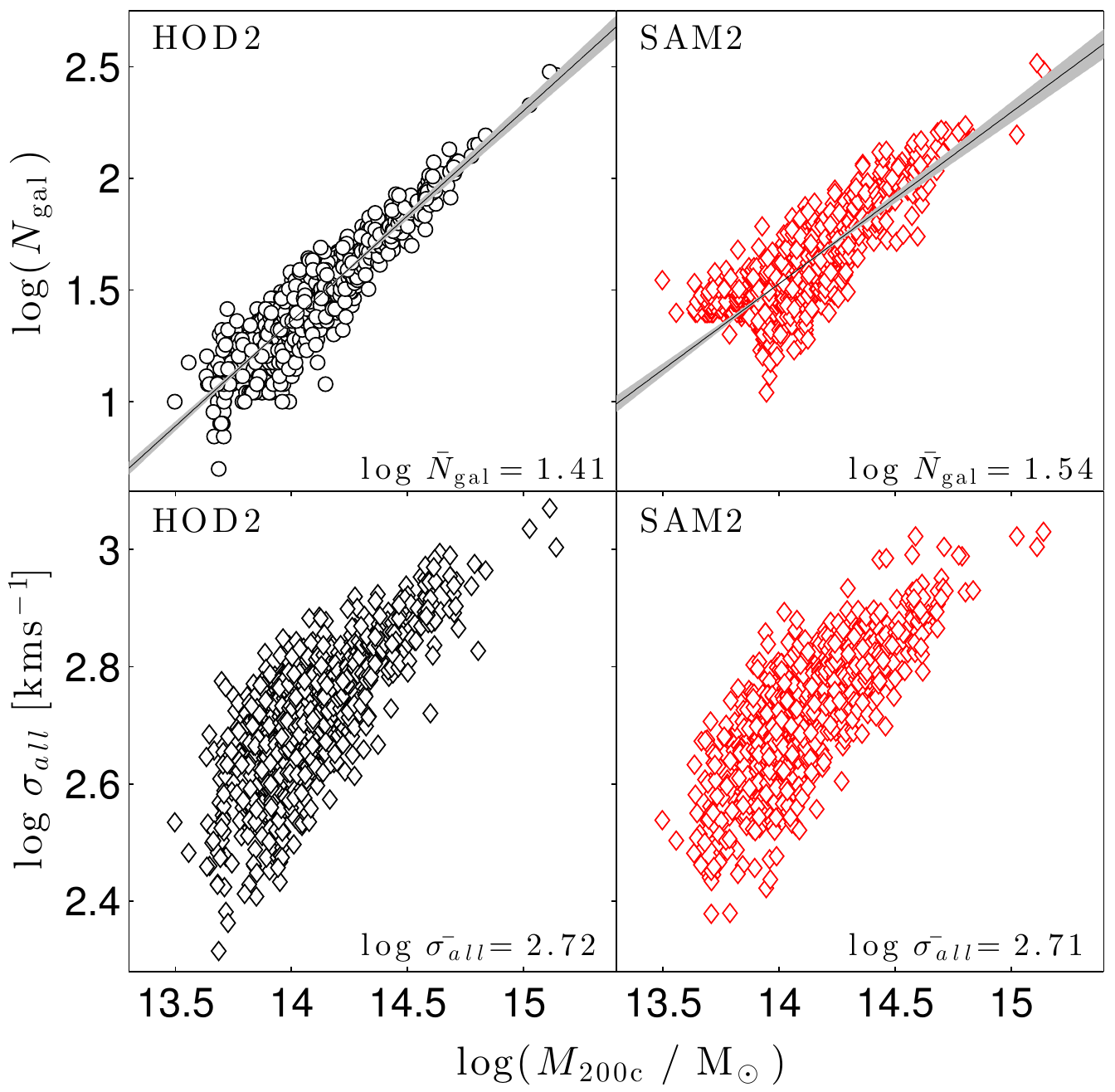}
 \caption{Upper row: richness versus mass of the 968 groups/clusters for both the HOD2
   and SAM2 input catalogues. The intrinsic scatter of the richness versus mass relation of the HOD2 and SAM2 catalogue is 0.09~dex and 0.12~dex, respectively. Lower row: velocity dispersion versus mass of the 968 groups/clusters for both the HOD2
   and SAM2 input catalogues. The velocity dispersion is calculated by taking the deviation of the line-of-sight velocities of all member galaxies associated with the groups/clusters. The intrinsic scatter of the velocity dispersion versus mass relation is 0.071~dex and 0.066~dex for HOD2 and SAM2, respectively. We note that this scatter is higher than \citet{2013MNRAS.430.2638M}, who compute the 3D velocity dispersion, while we consider the line-of-sight velocity dispersion, which naturally results in a larger scatter, especially if the clusters are triaxial (see discussion in \citet{2010A&A...520A..30M}.}
\label{fig:HOD2_SAM2_N_vs_M}
\end{figure*}

\begin{table*} 

%\appendix
\begin{flushleft}

\section*{Appendix C: Mass recovery accuracy for low and high group/cluster samples}
\end{flushleft}

\renewcommand\thetable{C1} 
\caption{Mass recovery accuracy (RMS, scatter in the recovered mass ($\sigma_{M_{\rm Rec}}$), slope, scatter about the true mass ($\sigma_{M_{\rm
      True}}$) and the ranking based on $\sigma_{M_{\rm True}}$) for low- and
  high-mass clusters (split according to the median true mass in each subsample) for the HOD2
  input catalogue.  
A method producing an RMS scatter of below 0.2~dex is assigned 8 stars and then decreasing numbers of stars are assigned in subsequent bins of size 0.05~dex. The final bin of methods producing an RMS scatter greater than 0.5~dex, is given one star}\begin{center}
\begin{tabular}{ l c c c c c l c c c c c l}
\hline
\multicolumn{1}{c}{Method}&
 \multicolumn{6}{c}{HOD2 low masses} &
  \multicolumn{6}{c}{HOD2 high masses}\\ [0.8ex] 
  \cmidrule(r){2-7} \cmidrule(l){8-13} 
&RMS (dex)&$\sigma_{M_{\rm Rec}}$&Slope&$\sigma_{M_{\rm True}}$&Bias&Merit &RMS (dex)&$\sigma_{M_{\rm Rec}}$&Slope&$\sigma_{M_{\rm True}}$&Bias&Merit\\  [1.0ex] 
\hline
PCN&0.23&0.23&1.21&0.19&$\phantom{-}0.03$&******* &0.29&0.18&1.28&0.14&$\phantom{-}0.14$&******  \\
PFN&0.21&0.21&0.79&0.27&$-0.01$&******* &0.19&0.18&0.94&0.19&$-0.04$&********\\
NUM&0.17&0.16&0.72&0.22&$-0.06$&********&0.19&0.13&0.82&0.16&$-0.09$&********\\
RM1&0.22&0.18&0.57&0.32&$\phantom{-}0.11$&******* &0.20&0.17&0.99&0.17&$\phantom{-}0.11$&********\\
RM2&0.23&0.20&0.55&0.37&$\phantom{-}0.09$&******* &0.20&0.17&1.00&0.17&$\phantom{-}0.11$&********\\
ESC&0.44&0.44&0.78&0.57&$-0.02$&***     &0.26&0.26&1.16&0.22&$-0.05$&******  \\
MPO&0.40&0.39&1.28&0.31&$-0.07$&****    &0.30&0.28&1.17&0.24&$-0.01$&*****   \\
MP1&0.41&0.34&1.07&0.32&$-0.20$&***     &0.33&0.23&1.07&0.22&$-0.17$&*****   \\
RW &0.37&0.35&1.24&0.28&$-0.10$&****    &0.29&0.27&1.07&0.25&$-0.11$&******  \\
TAR&0.31&0.27&1.01&0.27&$-0.14$&*****   &0.23&0.21&0.99&0.21&$-0.10$&******* \\
PCO&0.44&0.42&1.48&0.29&$\phantom{-}0.03$&***     &0.35&0.29&1.43&0.20&$\phantom{-}0.17$&*****   \\
PFO&0.47&0.40&1.60&0.25&$\phantom{-}0.12$&**      &0.37&0.27&1.31&0.21&$\phantom{-}0.20$&****    \\
PCR&1.28&0.97&0.93&1.04&$-0.84$&*       &0.82&0.56&0.99&0.56&$-0.60$&*       \\
PFR&0.49&0.42&0.45&0.94&$-0.26$&**      &0.53&0.34&0.63&0.54&$-0.39$&*       \\
MVM&0.19&0.16&0.59&0.26&$\phantom{-}0.08$&********&0.16&0.13&0.61&0.20&$-0.00$&********\\
AS1&0.50&0.49&1.10&0.45&$\phantom{-}0.11$&**      &0.38&0.37&1.02&0.37&$\phantom{-}0.08$&****    \\
AS2&0.53&0.49&0.99&0.49&$\phantom{-}0.21$&*       &0.40&0.38&0.92&0.41&$\phantom{-}0.15$&***     \\
AvL&0.38&0.34&1.09&0.32&$\phantom{-}0.16$&****    &0.29&0.25&1.08&0.23&$\phantom{-}0.15$&******  \\
CLE&0.43&0.42&1.08&0.39&$-0.09$&***     &0.33&0.30&1.11&0.27&$-0.14$&*****   \\
CLN&0.48&0.36&1.27&0.28&$-0.28$&**      &0.37&0.25&1.11&0.22&$-0.23$&****    \\
SG1&0.50&0.50&1.11&0.45&$\phantom{-}0.10$&*       &0.35&0.35&0.99&0.35&$\phantom{-}0.03$&*****   \\
SG2&0.45&0.36&0.97&0.37&$-0.14$&**      &0.32&0.24&1.00&0.24&$-0.18$&*****   \\
SG3&0.29&0.28&1.08&0.26&$-0.08$&******  &0.22&0.21&1.05&0.20&$-0.04$&******* \\
PCS&0.37&0.33&1.09&0.30&$-0.16$&****    &0.31&0.26&1.13&0.23&$-0.18$&*****   \\
PFS&0.37&0.34&1.42&0.24&$-0.15$&****    &0.33&0.27&1.14&0.24&$-0.15$&*****   \\
\hline
\end{tabular}
\end{center}
\label{table:HOD2_splitmasses_table}
\end{table*}
\begin{table*} 
\renewcommand\thetable{C2} 
\caption{Mass recovery accuracy for low and high mass groups/SAM2 clusters. Same notation as Table~\ref{table:HOD2_splitmasses_table}.}
%\begin{center}
%\begin{tabular}{ l c c c L L l c c c L L l}
%\hline
%\multicolumn{1}{c}{Method}&
% \multicolumn{6}{c}{SAM2 low masses} &
%  \multicolumn{6}{c}{SAM2 high masses}\\ [0.8ex] 
%  \cmidrule(r){2-7} \cmidrule(l){8-13} 
%&RMS (dex)&$\sigma_{M_{\rm Rec}}$&Slope&$\sigma_{M_{\rm True}}$&Bias&Merit
%  &RMS (dex)&$\sigma_{M_{\rm Rec}}$&Slope&$\sigma_{M_{\rm True}}$&Bias&Merit\\  [1.0ex] 
%  \hline
\begin{center}
\begin{tabular}{ l c c c c c l c c c c c l}
\hline
\multicolumn{1}{c}{Method}&
 \multicolumn{6}{c}{SAM2 low masses} &
  \multicolumn{6}{c}{SAM2 high masses}\\ [0.8ex] 
  \cmidrule(r){2-7} \cmidrule(l){8-13} 
&RMS (dex)&$\sigma_{M_{\rm Rec}}$&Slope&$\sigma_{M_{\rm True}}$&Bias&Merit
  &RMS (dex)&$\sigma_{M_{\rm Rec}}$&Slope&$\sigma_{M_{\rm True}}$&Bias&Merit\\  [1.0ex] 
  \hline
PCN&0.41&0.23& 0.12&$\phantom{-}1.96$&$\phantom{-}0.31$&***     &0.35&0.22&0.90&$\phantom{-}0.25$&$\phantom{-}0.27$&*****   \\
PFN&0.51&0.37&0.26&$\phantom{-}1.41$&$\phantom{-}0.34$&*       &0.47&0.38&0.97&$\phantom{-}0.40$&$\phantom{-}0.28$&**      \\
NUM&0.19&0.15&0.03&$-5.41$&$\phantom{-} 0.03$&********&0.22&0.15&0.50&$\phantom{-}0.30$&$-0.10$&******* \\
ESC&0.42&0.41& 1.13&$\phantom{-} 0.36$&$-0.02$&***     &0.38&0.26&1.08&$\phantom{-}0.24$&$-0.01$&****    \\
MPO&0.31&0.31& 1.01&$\phantom{-} 0.30$&$-0.01$&*****   &0.23&0.22&0.89&$\phantom{-}0.25$&$-0.04$&******* \\
MP1&0.30&0.26& 0.79&$\phantom{-} 0.33$&$-0.14$&******  &0.32&0.21&0.71&$\phantom{-}0.29$&$-0.22$&*****   \\
RW &0.34&0.33& 0.89&$\phantom{-} 0.37$&$\phantom{-}0.06$&*****   &0.26&0.26&0.95&$\phantom{-}0.27$&$\phantom{-} 0.04$&******  \\
TAR&0.30&0.30& 0.37&$\phantom{-} 0.79$&$-0.03$&*****   &0.31&0.31&1.09&$\phantom{-}0.28$&$-0.06$&*****   \\
PCO&0.41&0.38&0.06&$-6.36$&$\phantom{-} 0.11$&***     &0.41&0.38&1.30&$\phantom{-}0.29$&$\phantom{-} 0.09$&***     \\
PFO&0.60&0.46& 0.13&$\phantom{-} 3.53$&$\phantom{-} 0.20$&*       &0.64&0.57&1.49&$\phantom{-}0.38$&$\phantom{-} 0.13$&*       \\
PCR&0.71&0.54& 0.70&$\phantom{-}0.77$&$\phantom{-} 0.45$&*       &0.56&0.37&0.60&$\phantom{-}0.62$&$\phantom{-} 0.43$&*       \\
PFR&0.67&0.50& 0.37&$\phantom{-}1.35$&$\phantom{-} 0.42$&*       &0.57&0.46&0.75&$\phantom{-}0.60$&$\phantom{-} 0.35$&*       \\
MVM&0.33&0.14& 0.74&$\phantom{-}0.19$&$\phantom{-} 0.29$&*****   &0.22&0.12&0.55&$\phantom{-}0.22$&$\phantom{-} 0.19$&******* \\
AS1&0.63&0.62& 1.23&$\phantom{-}0.51$&$-0.08$&*       &0.45&0.45&1.30&$\phantom{-}0.34$&$-0.04$&***     \\
AS2&0.62&0.62& 1.18&$\phantom{-}0.52$&$\phantom{-}0.06$&*       &0.44&0.44&1.20&$\phantom{-}0.36$&$\phantom{-}0.07$&***     \\
AvL&0.35&0.30& 1.27&$\phantom{-}0.24$&$\phantom{-} 0.19$&****    &0.31&0.24&1.12&$\phantom{-}0.22$&$\phantom{-} 0.19$&*****   \\
CLE&0.33&0.31& 1.32&$\phantom{-} 0.24$&$-0.11$&*****   &0.28&0.25&1.07&$\phantom{-}0.24$&$-0.11$&******  \\
CLN&0.37&0.30& 1.14&$\phantom{-} 0.26$&$-0.18$&****    &0.31&0.24&1.00&$\phantom{-}0.24$&$-0.19$&*****   \\
SG1&0.46&0.45& 0.96&$\phantom{-} 0.47$& $\phantom{-}0.11$&**      &0.33&0.32&0.95&$\phantom{-}0.34$& $\phantom{-}0.08$&*****   \\
SG2&0.33&0.32& 1.26&$\phantom{-} 0.25$&$-0.07$&*****   &0.29&0.24&1.01&$\phantom{-}0.24$&$-0.12$&******  \\
SG3&0.30&0.21& 0.89&$\phantom{-} 0.24$& $\phantom{-}0.21$&*****   &0.26&0.18&0.93&$\phantom{-}0.19$& $\phantom{-}0.20$&******  \\
PCS&0.36&0.31& 1.23&$\phantom{-} 0.25$&$-0.18$&****    &0.28&0.25&1.27&$\phantom{-}0.20$&$-0.14$&******  \\
PFS&0.56&0.49& 0.78&$\phantom{-} 0.63$&$-0.28$&*       &0.55&0.44&1.17&$\phantom{-}0.38$&$-0.31$&*       \\
\hline
\end{tabular}
\end{center}
\label{table:SAM2_splitmasses_table}
\end{table*}

\clearpage
\begin{table*} 
\begin{flushleft}
\section*{Appendix D: Richness recovery}
\end{flushleft}
\renewcommand\thetable{D1} 
\caption{The RMS, scatter in the observed richness, $\sigma_{N_{\rm gal, obs}}$, slope, scatter about the true richness $\sigma_{N_{\rm gal, true}}$ and the bias at the pivot richness (for the HOD2: log $N_{\rm gal, true}=1.41$ and for the SAM2: log $N_{\rm gal, true}=1.54$).}
\begin{center}
%\begin{tabular}{p{0.35cm}p{1.25cm}P{0.95cm}P{0.95cm}P{0.95cm}p{0.95cm}P{1.25cm}P{0.95cm}P{0.95cm}P{0.95cm}P{0.95cm}}
\begin{tabular}{ l c c c c c c c c c c}
\hline
\multicolumn{1}{c}{Method}&
 \multicolumn{5}{c}{HOD2} &
  \multicolumn{5}{c}{SAM2}\\ [0.8ex] 
  \cmidrule(r){2-6} \cmidrule(l){7-11} 
&RMS (dex)&$\sigma_{N_{\rm gal, obs}}$&Slope&$\sigma_{N_{\rm gal, true}}$& Bias &RMS (dex)&$\sigma_{N_{\rm gal, obs}}$&Slope&$\sigma_{N_{\rm gal, true}}$& Bias\\  [1.0ex] 
\hline
PCN&0.12&0.09&0.86&0.10&$\phantom{-}0.00$&0.15&0.12&0.62&0.19&$-0.00$\\
PFN&0.19&0.17&0.92&0.18&$\phantom{-}0.09 $&0.44&0.35&0.89&0.39&$\phantom{-}0.27 $\\
NUM&0.11&0.10&0.97&0.10&$\phantom{-}0.02 $&0.16&0.15&0.77&0.19&$\phantom{-}0.04 $\\
RM1&0.17&0.14&0.95&0.15&$-0.08$&&&&\\
RM2&0.18&0.15&0.98&0.15&$-0.09$&&&&&\\
ESC&0.14&0.14&0.94&0.14&$\phantom{-}0.03$&0.25&0.20&0.90&0.23&$\phantom{-}0.03$\\
MPO&0.13&0.12&1.05&0.12&$-0.04$&0.20&0.19&0.89&0.21&$-0.07$\\
MP1&0.15&0.13&1.03&0.13&$-0.07$&0.23&0.19&0.84&0.23&$-0.11$\\
RW &0.13&0.13&0.98&0.13&$\phantom{-}0.00$&0.18&0.18&0.88&0.20&$\phantom{-}0.02$\\
TAR&0.13&0.13&0.94&0.14&$\phantom{-}0.01$&0.24&0.24&0.90&0.26&$\phantom{-}0.04$\\
PCO&0.10&0.09&0.85&0.11&$\phantom{-}0.01$&0.14&0.12&0.61&0.19&$\phantom{-}0.00$\\
PFO&0.19&0.16&0.90&0.18&$\phantom{-}0.10$&0.43&0.32&0.91&0.35&$\phantom{-}0.29$\\
PCR&0.12&0.09&0.86&0.10&$\phantom{-}0.00$&0.15&0.12&0.62&0.19&$-0.00$\\
PFR&0.19&0.17&0.92&0.18&$\phantom{-}0.09$&0.44&0.35&0.89&0.39&$\phantom{-}0.27$\\
MVM&0.11&0.09&0.77&0.11&$\phantom{-}0.02$&0.17&0.12&0.63&0.18&$\phantom{-}0.12$\\
AS1&0.22&0.17&0.78&0.22&$-0.13$&0.54&0.15&0.56&0.27&$-0.48$\\
AS2&0.22&0.17&0.78&0.22&$-0.13$&0.54&0.15&0.56&0.27&$-0.48$\\
AvL&0.12&0.11&0.97&0.11&$\phantom{-}0.01$&0.17&0.17&0.92&0.19&$\phantom{-}0.01$\\
CLE&0.15&0.15&0.99&0.15&$\phantom{-}0.01$&0.20&0.20&0.93&0.21&$-0.02$\\
CLN&0.14&0.13&1.07&0.12&$-0.04$&0.21&0.19&0.90&0.21&$-0.07$\\
SG1&0.18&0.17&0.94&0.18&$\phantom{-}0.05$&0.23&0.22&0.91&0.25&$\phantom{-}0.08$\\
SG2&0.17&0.16&0.97&0.16&$-0.04$&0.22&0.20&0.92&0.22&$-0.08$\\
SG3&0.21&0.18&0.99&0.18&$\phantom{-}0.12$&0.39&0.16&0.69&0.22&$\phantom{-}0.37$\\
PCS&0.12&0.09&0.86&0.10&$\phantom{-}0.00$&0.15&0.12&0.62&0.19&$-0.00$\\
PFS&0.19&0.17&0.92&0.18&$\phantom{-}0.09$&0.44&0.35&0.90&0.39&$\phantom{-}0.27$\\
\hline
\end{tabular}
\vspace{1.5cm}
\end{center}
\label{table:HOD2_SAM2_Ngal_table}
\end{table*}
\clearpage
\begin{figure*}
\begin{flushleft}
\section*{Appendix E: Residuals and recovered mass distributions}
\end{flushleft}
\renewcommand\thefigure{E1} 
 \centering
 \includegraphics[trim = 45mm 52mm 55mm 57mm, clip, width=0.80\textwidth]{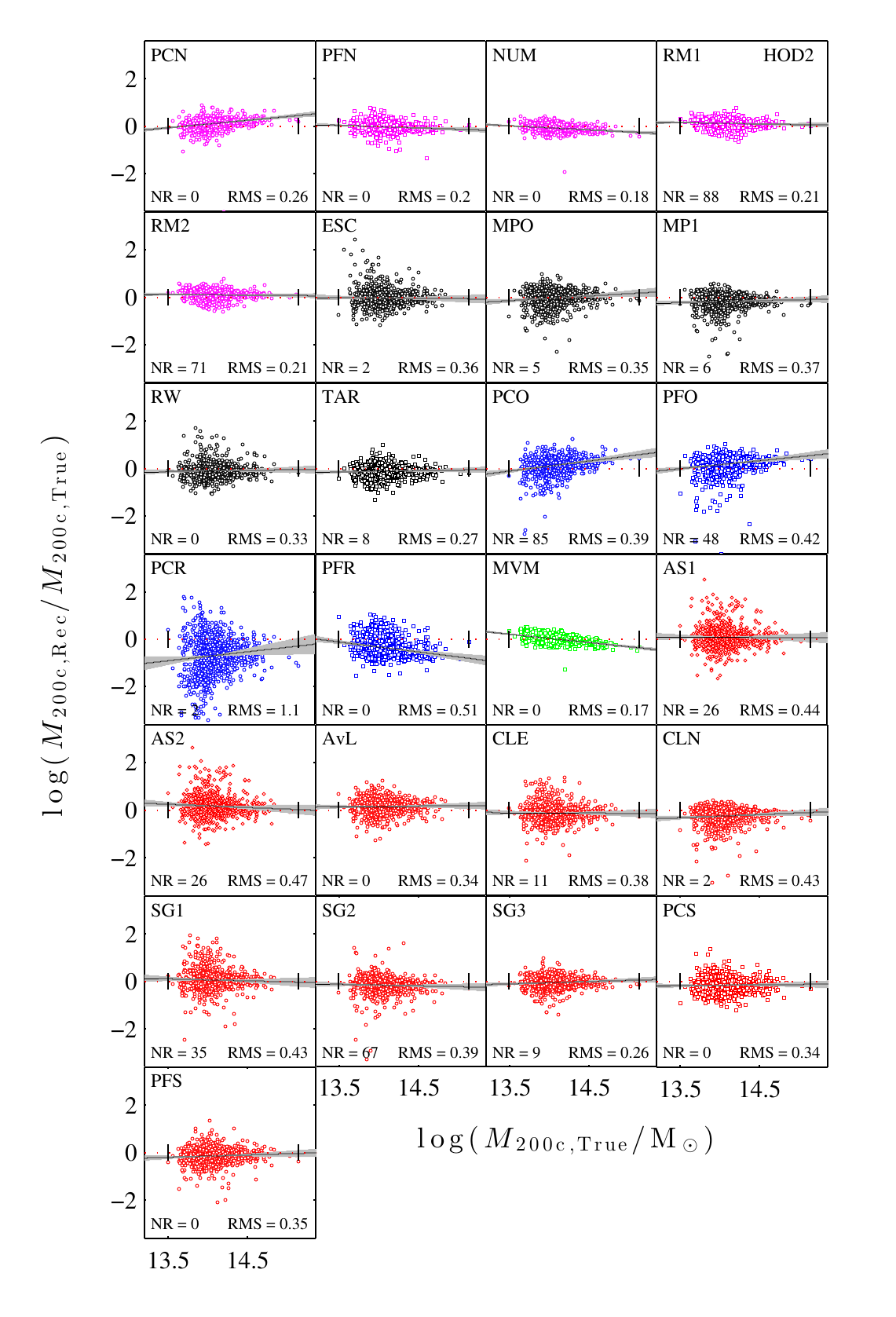}
 \caption{Residuals of the recovered versus true cluster mass for the 25
   methods using the HOD2 input
   catalogue. This Figure follows the same notation as in Figure~\ref{fig:HOD2_SW_Nsigma_outliers_mass_scatter_combined}.}
\label{fig:HOD2_SW_residual_mass_scatter_combined}
\end{figure*}

\clearpage
\begin{figure*}
\renewcommand\thefigure{E2} 
 \centering
\includegraphics[trim = 40mm 43mm 50mm 55mm, clip, width=0.83\textwidth]{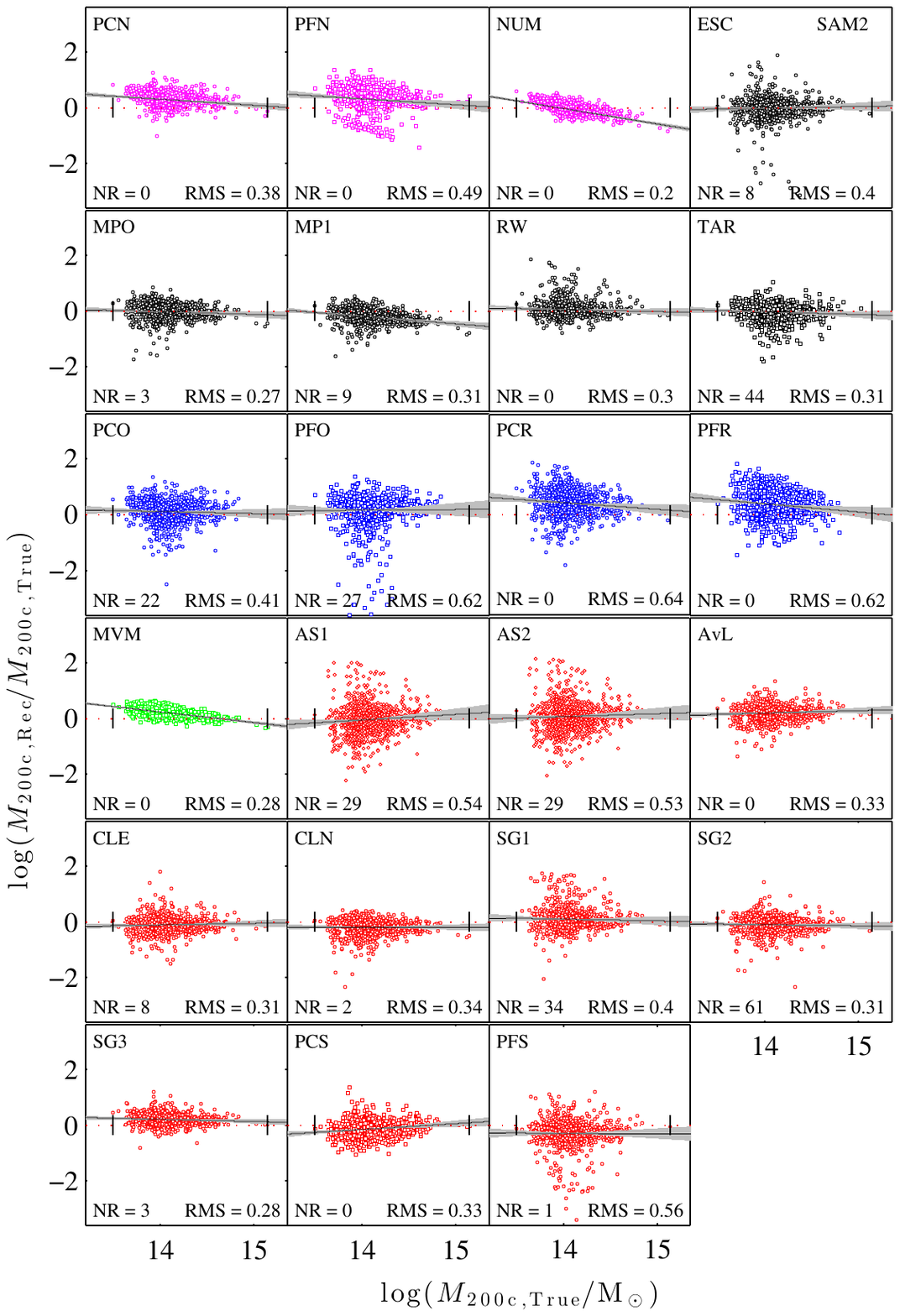}
 \caption{Residuals of the recovered versus true cluster mass for the 23
   methods applied to the SAM2 input
   catalogue. This Figure follows the same notation as in Figure~\ref{fig:HOD2_SW_residual_mass_scatter_combined}.}
\label{fig:SAM2_SW_residual_mass_scatter_combined}
\end{figure*}

\clearpage

\begin{figure*}
\renewcommand\thefigure{E3} 
 \centering
\includegraphics[trim = 45mm 52mm 55mm 57mm, clip, width=0.80\textwidth]{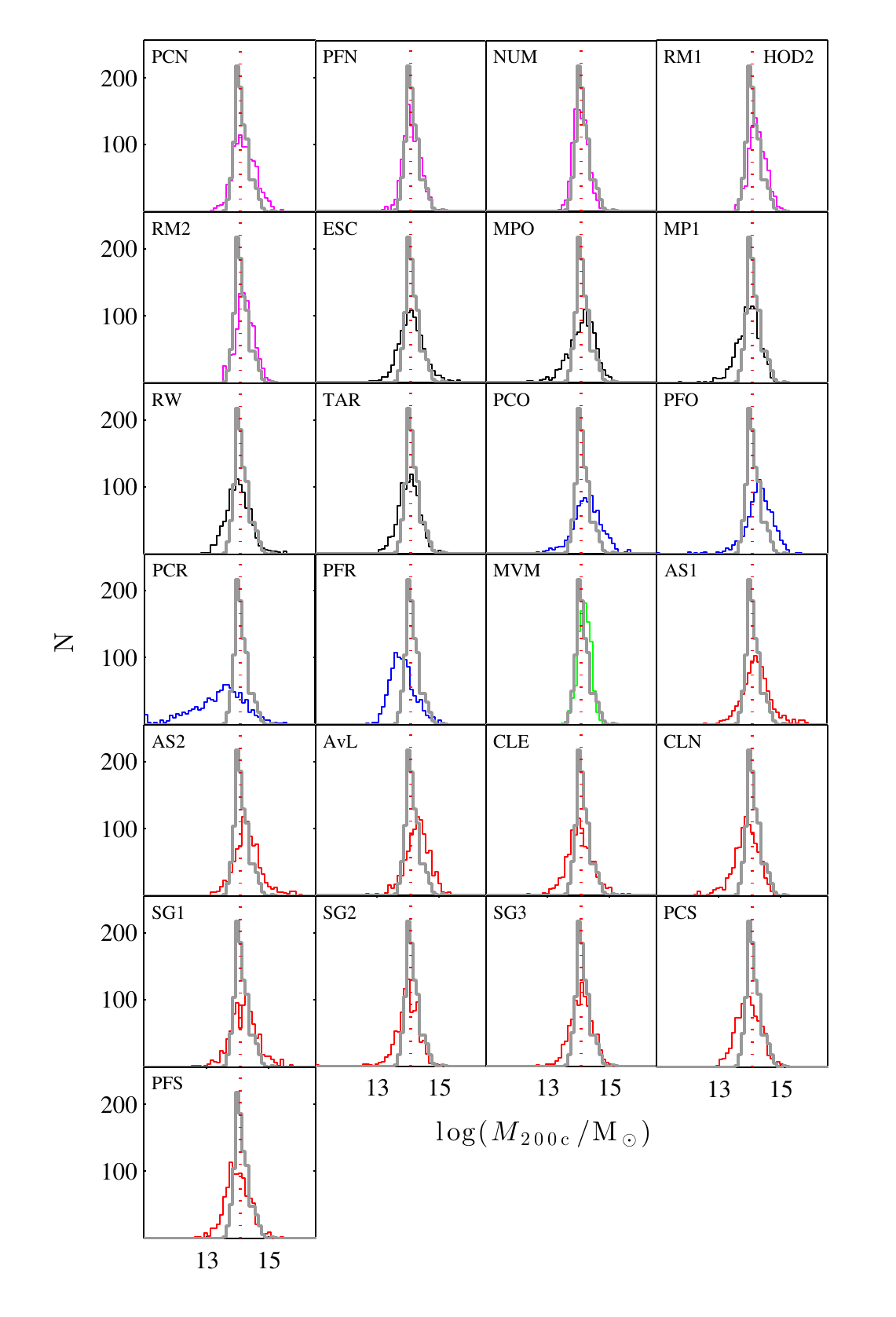}
 \caption{Recovered cluster mass distributions for the 25 methods applied to the HOD2
   input catalogue. The red dotted line represents the mean of
 the true mass distribution and the grey distributions on each
 subplot represent the true mass distributions.}
\label{fig:HOD2_SW_mass_histograms_combined}
\end{figure*}

\clearpage

\begin{figure*}
\renewcommand\thefigure{E4} 
 \centering
\includegraphics[trim = 40mm 43mm 50mm 55mm, clip, width=0.83\textwidth]{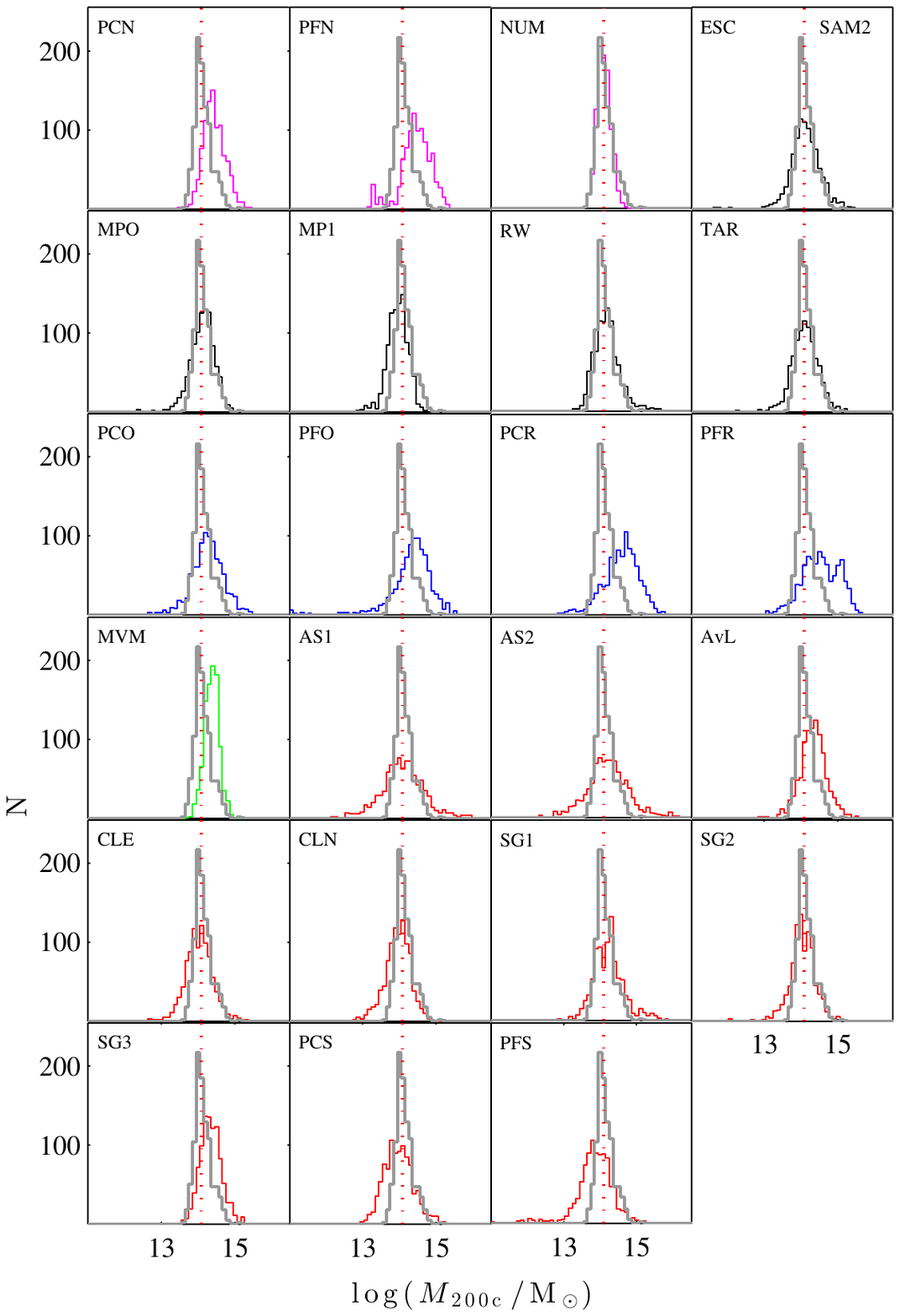}
 \caption{Recovered mass distributions for the 23 methods applied to the SAM2
   input catalogue. This Figure follows the same notation as in Figure~\ref{fig:HOD2_SW_mass_histograms_combined}.}
\label{fig:SAM2_SW_mass_histograms_combined}
\end{figure*}

\label{lastpage}

\end{document}